\documentclass{aa}  

\usepackage{graphicx}
\usepackage{txfonts}
\usepackage{hyperref}
\usepackage{natbib}

\hypersetup{
        colorlinks=true,    
        linkcolor=blue,  
        citecolor=blue,    
        filecolor=cyan,     
        urlcolor=black,      
} 

\begin{document} 

\title{Massive galaxy formation caught in action at $z\sim5$  with JWST}


\titlerunning{JWST reveals a compact galaxy group at $z\sim5.2$}
\authorrunning{Jin et al.}

\author{
Shuowen Jin\inst{1,2,\thanks{Marie Curie Fellow}},
Nikolaj B. Sillassen\inst{1,2},
Georgios E. Magdis\inst{1,2,3},
Aswin P. Vijayan\inst{1,2},
Gabriel B. Brammer\inst{1,3},
Vasily Kokorev\inst{4},
John~R.~Weaver\inst{5},
Raphael Gobat\inst{6},
Clara Gim\'enez-Arteaga\inst{1,3},
Francesco Valentino\inst{1,3},
Malte Brinch\inst{1,2},
Carlos G\'omez-Guijarro\inst{7},
Marko Shuntov\inst{1,3,8},
Sune Toft\inst{1,3},
Thomas R. Greve\inst{1,2},
and David Blanquez Sese\inst{1,2}
          }

   \institute{Cosmic Dawn Center (DAWN), Denmark\\
      \email{shuji@space.dtu.dk, shuowen.jin@gmail.com}
    \and
            DTU Space, Technical University of Denmark, Elektrovej 327, DK-2800 Kgs. Lyngby, Denmark
    \and
            Niels Bohr Institute, University of Copenhagen, Jagtvej 128, DK-2200 Copenhagen, Denmark
    \and
    Kapteyn Astronomical Institute, University of Groningen, PO Box 800, 9700 AV Groningen, The Netherlands
    \and
            Department of Astronomy, University of Massachusetts, Amherst, MA 01003, USA
    \and
            Instituto de Física, Pontificia Universidad Católica de Valparaíso, Casilla 4059, Valparaíso, Chile
    \and
            Universit\'e Paris-Saclay, Universit\'e Paris Cit\'e, CEA, CNRS, AIM, 91191, Gif-sur-Yvette, France
    \and
            Institut d'Astrophysique de Paris, UMR 7095, CNRS, and Sorbonne Universit\'e, 98 bis boulevard Arago, 75014 Paris, France
             }

   \date{Received XXX / Accepted XXX}

 \abstract
{
We report the discovery of a compact group of galaxies, CGG-z5, at $z\sim5.2$ in the EGS field covered by the JWST/CEERS survey. CGG-z5 was selected as the highest overdensity of galaxies at $z>2$ in recent JWST public surveys and it consists of six candidate members lying within a projected area of $1.5''\times3''$ (10$\times$20~kpc$^2$). All group members  are HST/F435W and HST/F606W dropouts while securely detected in the JWST/NIRCam bands, yielding a narrow range of robust photometric redshifts $5.0<z<5.3$. The most massive galaxy in the group has a stellar mass log$(M_{*}/M_{\odot})\approx9.8$, while the rest are low-mass satellites (log$(M_{*}/M_{\odot})\approx8.4$--9.2). While several group members were already detected in the HST and IRAC bands, the low stellar masses and the compactness of the structure required the sensitivity and resolution of JWST for its identification. To assess the nature and evolutionary path of CGG-z5, we searched for similar compact structures in the \textsc{Eagle} simulations and followed their evolution with time. We find that all the identified structures merge into a single galaxy by $z=3$ and form a massive galaxy  (log$(M_{*}/M_{\odot})>$11) at $z\sim1$. This implies that CGG-z5 could be a ``proto-massive galaxy''  captured during a short-lived phase of massive galaxy formation.}


\maketitle
 

\section{Introduction}

Understanding the formation of massive galaxies ($M_*>10^{11}~M_\odot$) is a major goal in modern astrophysics. 
In $\Lambda$CDM cosmology, massive galaxies are formed via hierarchical clustering, where massive dark matter haloes are formed through the continuous merging of low-mass subhaloes (e.g., \citealt{Springel2005Natur,Springel2008halo,Boylan-Kolchin2009,Klypin2011halo,Somerville_Dave2015}). 
The assembly of dark matter haloes, as one of the most important processes in the formation of structure in the Universe, can be traced by merger graphs or trees, based on analytical approaches and simulations (e.g., \citealt{Press_Schechter1974,Somerville1999,Zhang2008MergerTree,Neistein_Dekel2008,Roper2020}). 
Advanced cosmological simulations have displayed great details on this process, showing that the most rapid growth of nearby massive galaxies occurred at $z>2$ via the merging of low-mass satellites with $M_*\lesssim10^{10}~M_\odot$ as well as in situ gas accretion (e.g., \citealt{Hopkins2009,Oser2010simu,Benson2012MergerTree,Hirschmann2012simu}).

However, the observational confirmation of these merging systems before coalescence is still lacking, especially for low-mass systems at high redshift. 
This is not only due to the extreme faintness of low-mass galaxies at high redshift, but also their optical lines and stellar continuum are redshifted to infrared wavelengths that are not available with the Hubble Space Telescope (HST) and ground-based facilities. 
Now the situation has changed with the successful launch of the James Webb Space Telescope (JWST). With the unprecedented sensitivity, spatial resolution, and long wavelength coverage, we are right on the cusp of discovering more distant structures down to a significantly low-mass limit (e.g., \citealt{Morishita2022z8cluster}). 

In this Letter we report a low-mass and compact galaxy group CGG-z5 at $z\sim5.2$ revealed by JWST, a candidate of a merging system forming a massive galaxy. We adopt cosmology $H_0=73$~km~s$^{-1}$~Mpc$^{-1}$, $\Omega_M=0.27$, and  $\Lambda_0=0.73$ as well as a Chabrier initial mass function \citep{Chabrier2003}.

\begin{figure*}[ht]
\setlength{\abovecaptionskip}{-0.1cm}
\setlength{\belowcaptionskip}{-0.2cm}
\centering
\includegraphics[width=0.95\textwidth]{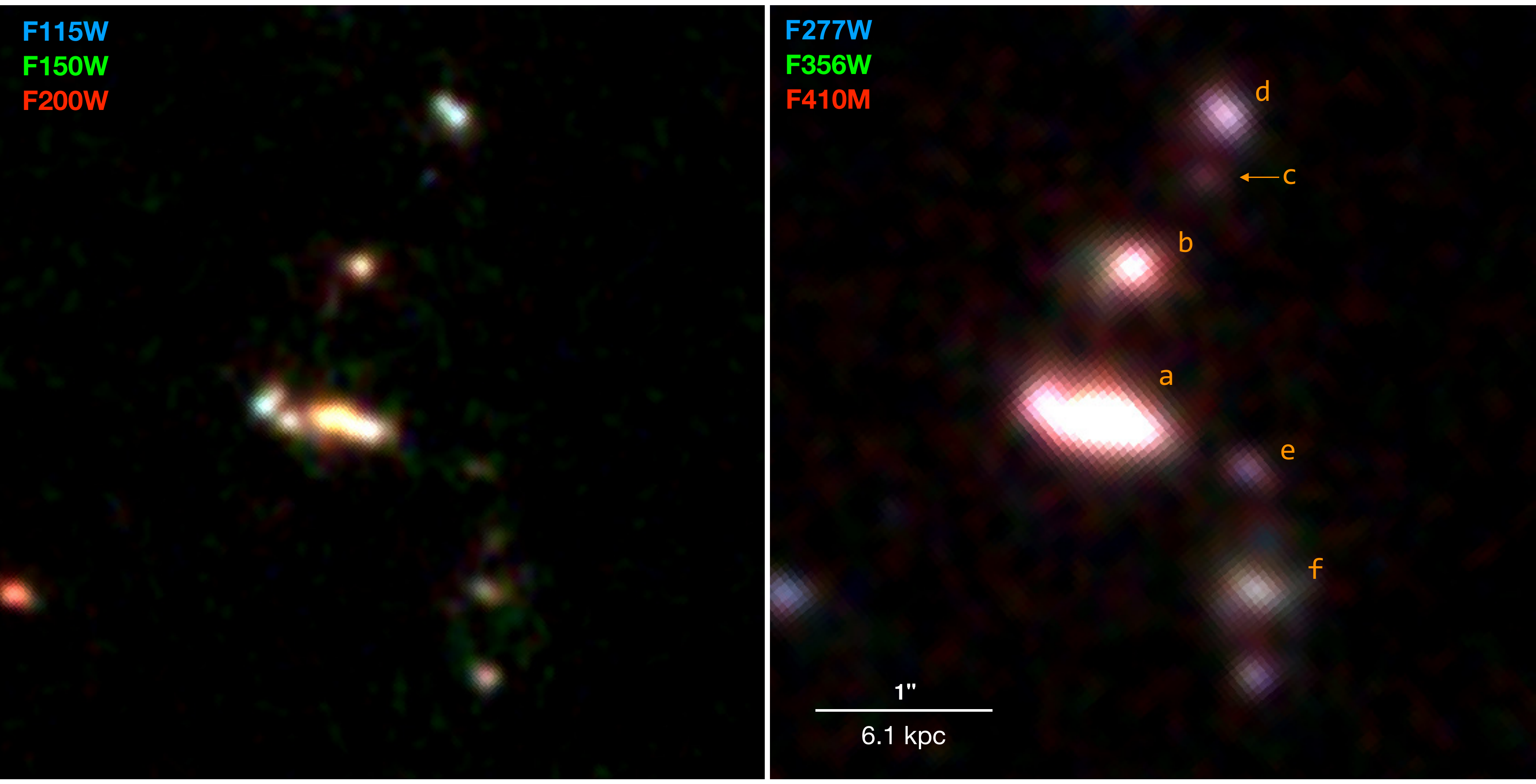}
\caption{%
JWST RGB color images of CGG-z5 at $z\sim5.2$. {\it Left:} Color image composed by NIRCam F115W (blue), F150W (green), and F200W (red) data.
{\it Right:} Color image composed by NIRCam F277W (blue), F356W (green), and F410M (red) data. Candidate members are labeled with text in orange.
The two color images are matched in the World Coordinate System, and the RGB frames are composed in linear scale with identical limits.
\label{img}
}
\end{figure*}

\begin{figure*}[ht]
\setlength{\abovecaptionskip}{-0.1cm}
\setlength{\belowcaptionskip}{-0.2cm}
\centering
\includegraphics[width=0.95\textwidth]{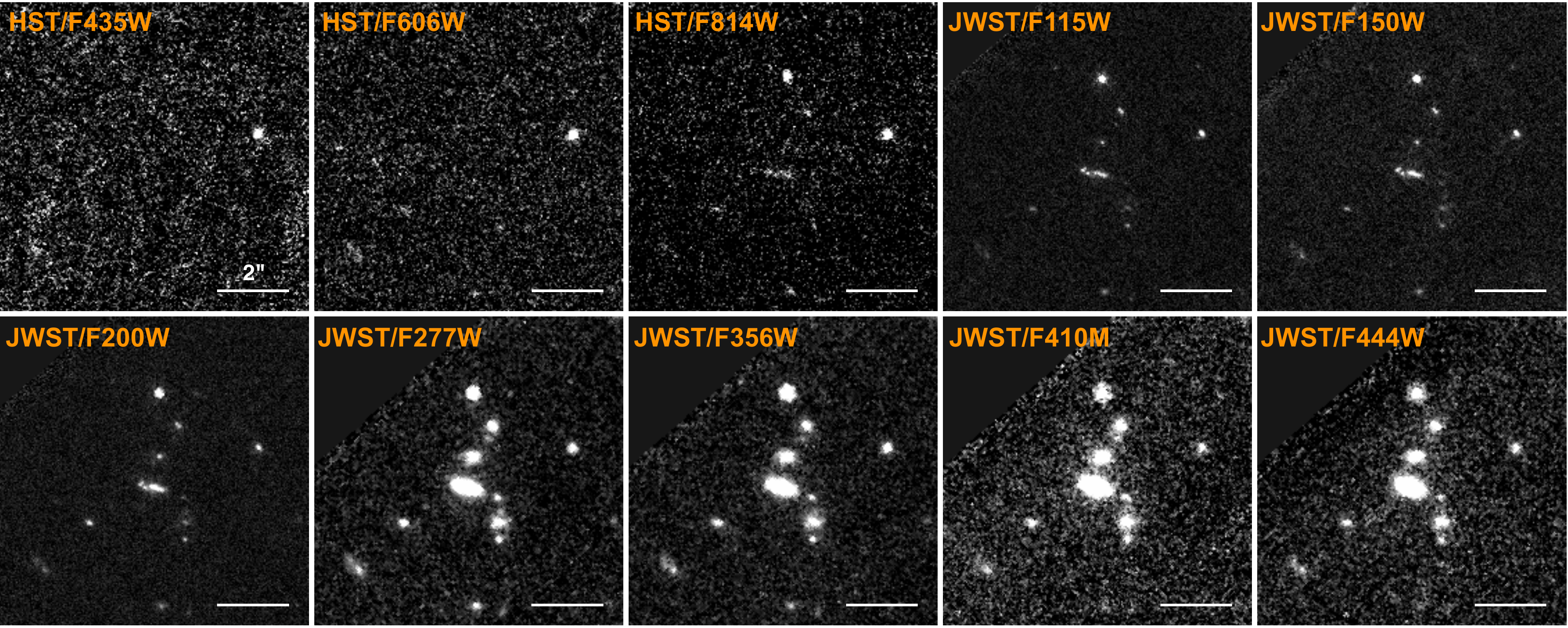}
\caption{%
HST and JWST cutouts in $8.5''\times8.5''$ size on CGG-z5. Telescopes and filters are shown in text in each cutout. The white bar in the bottom right marks a $2''$ scale (12 kpc). All images are shown in linear scale with identical limits.
\label{cutout}
}
\end{figure*}

\begin{figure*}[ht]
\setlength{\abovecaptionskip}{-0.1cm}
\setlength{\belowcaptionskip}{-0.2cm}
\centering
\includegraphics[width=0.45\textwidth]{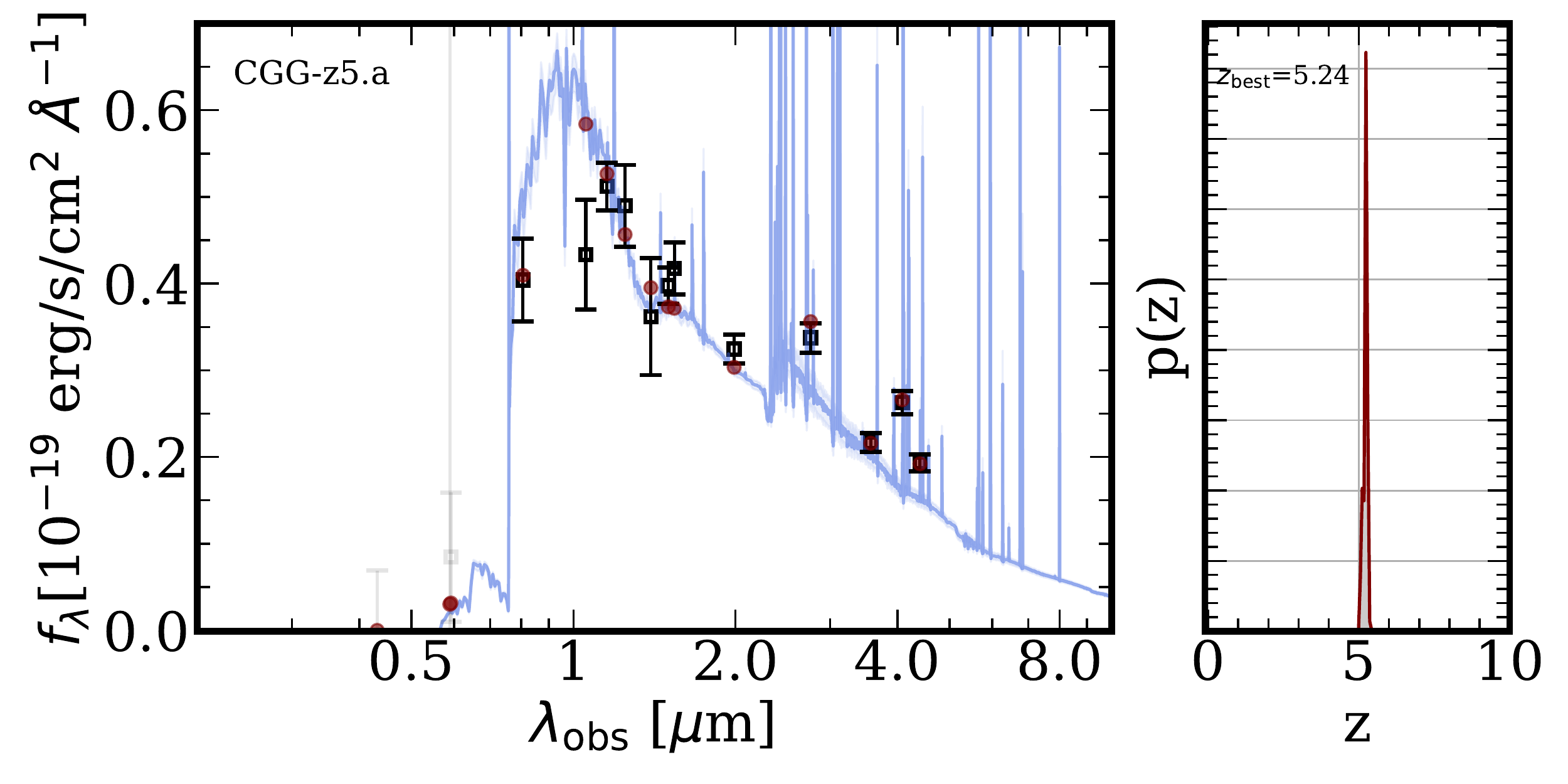}
\includegraphics[width=0.45\textwidth]{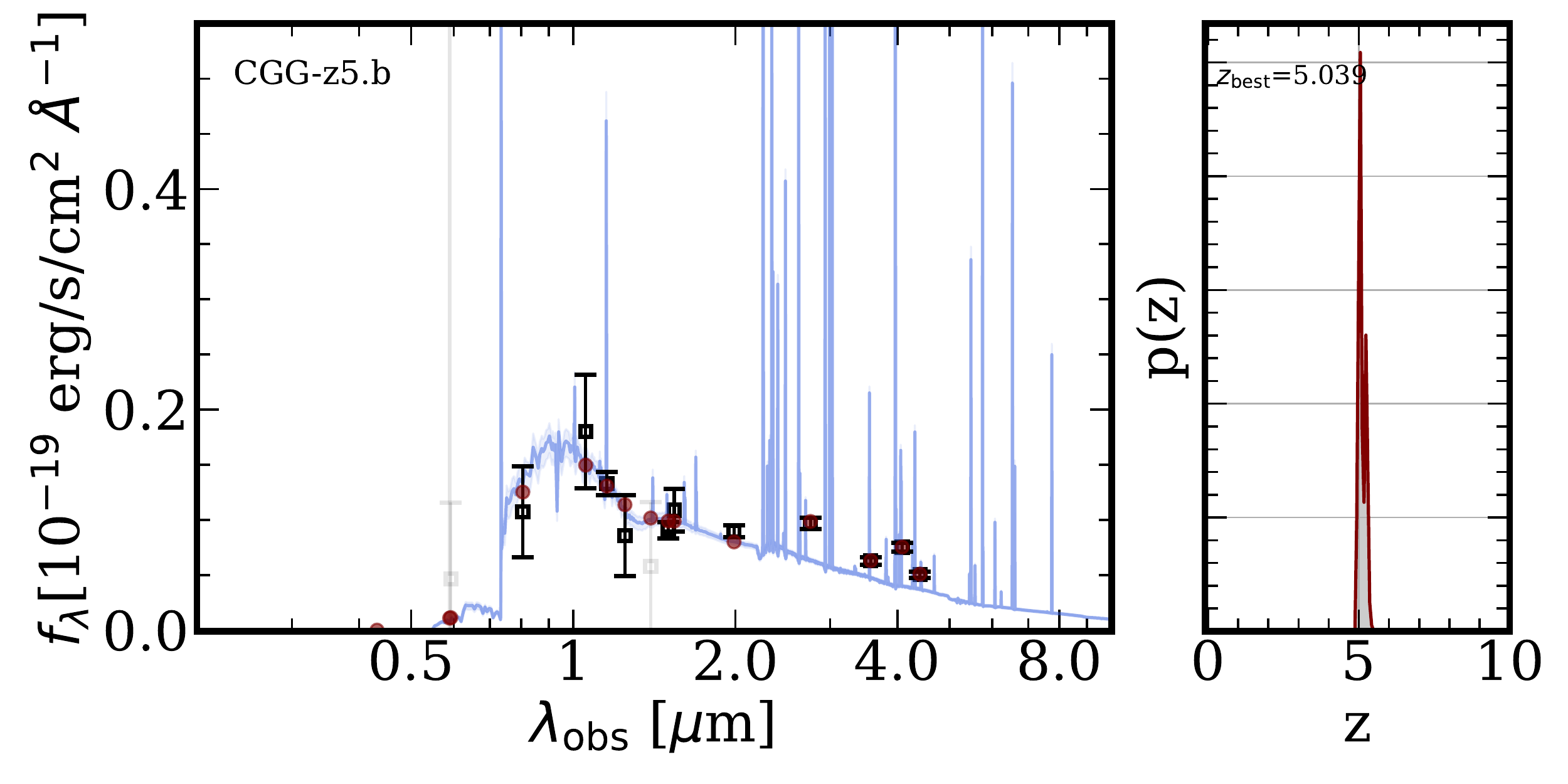}
\includegraphics[width=0.45\textwidth]{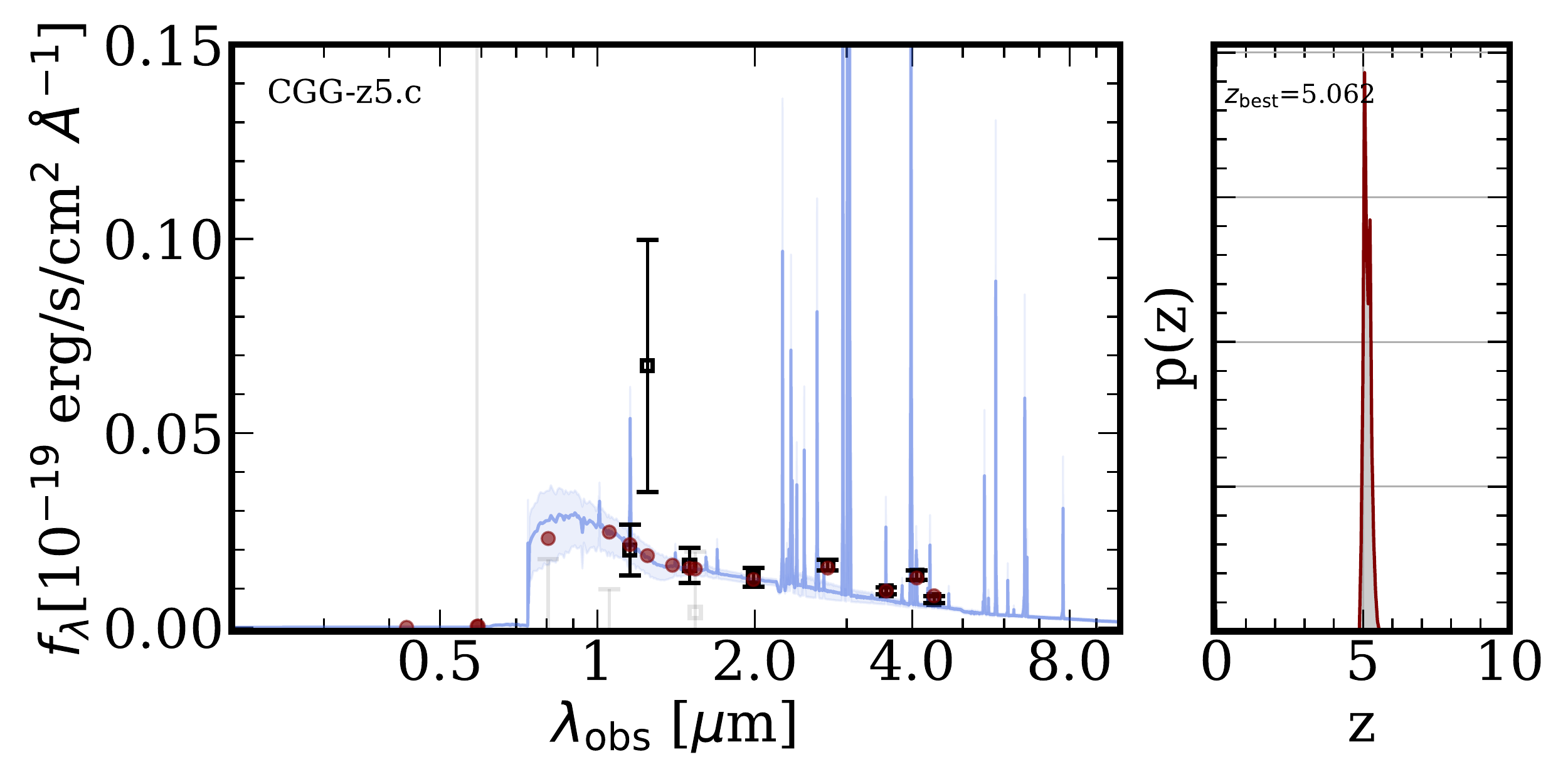}
\includegraphics[width=0.45\textwidth]{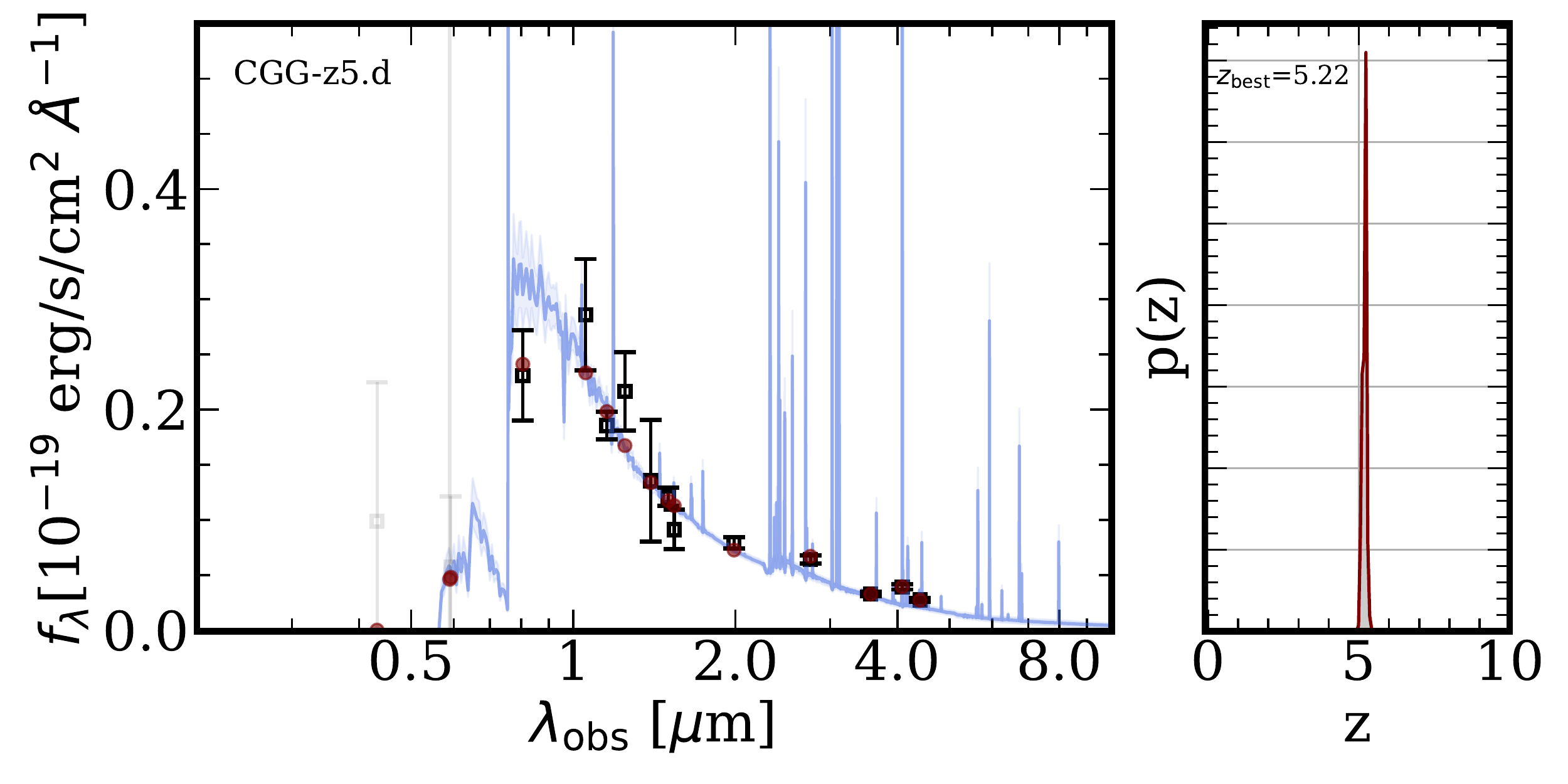}
\includegraphics[width=0.45\textwidth]{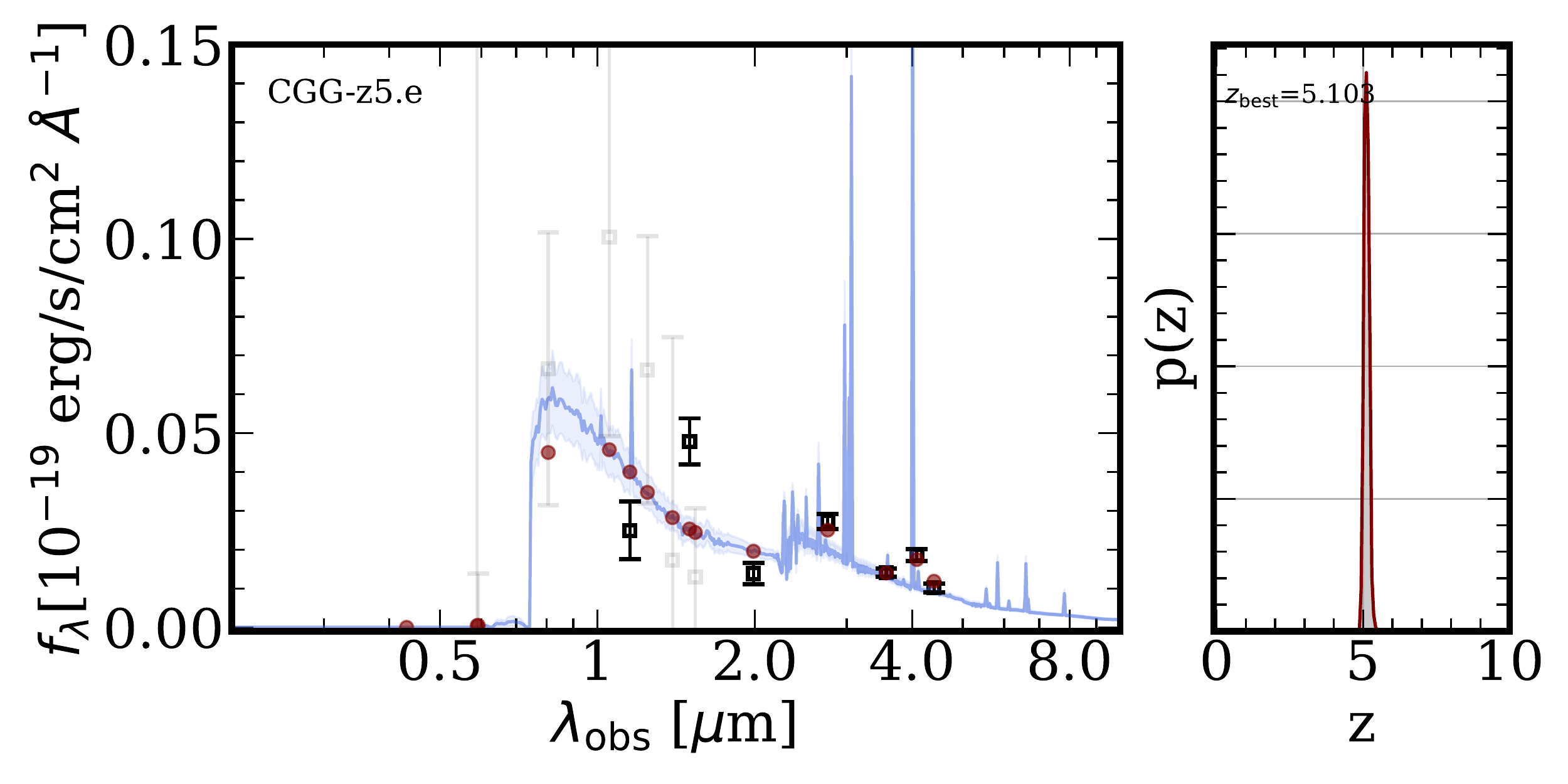}
\includegraphics[width=0.45\textwidth]{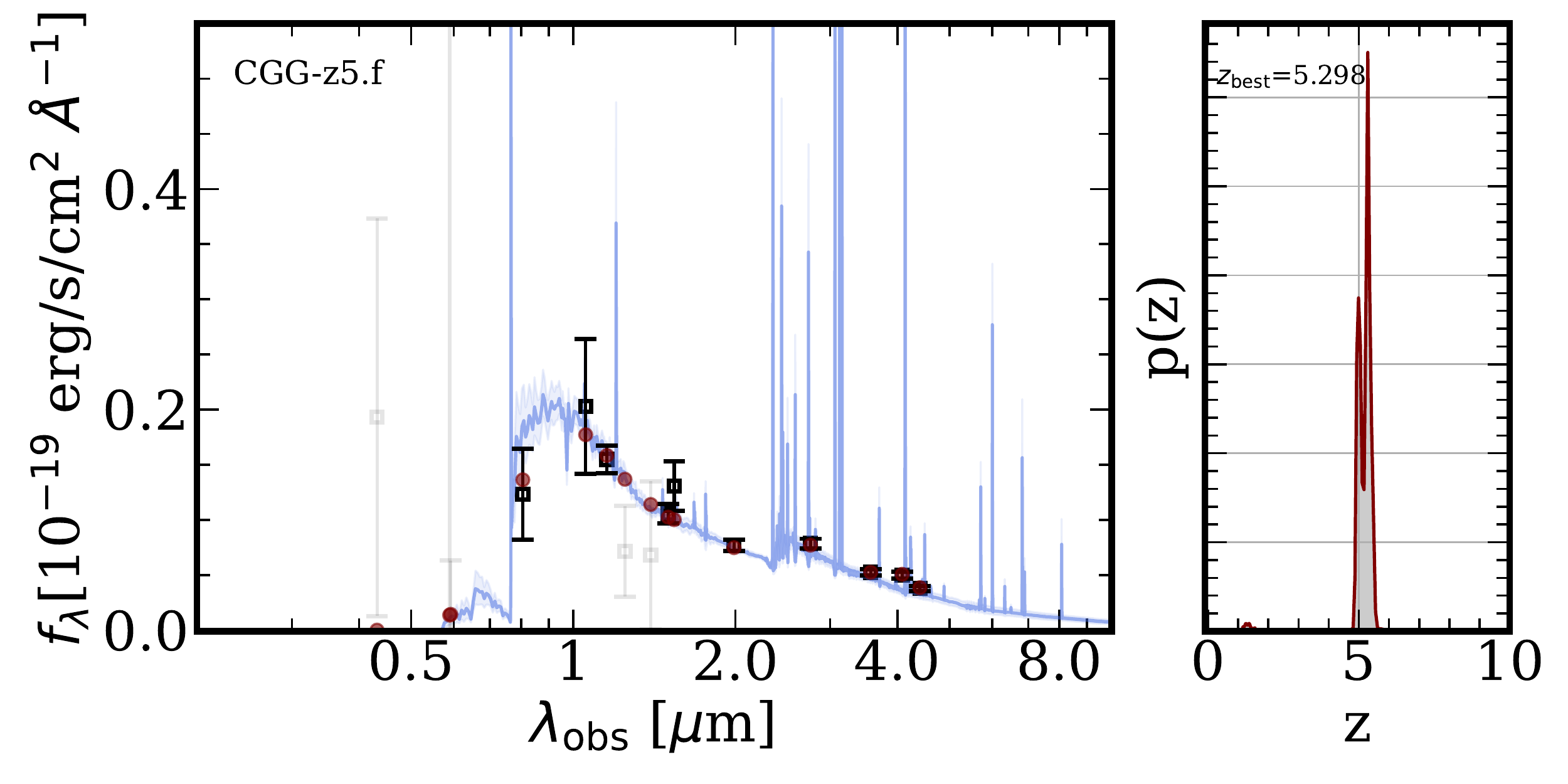}
\caption{%
SEDs of candidate members in CGG-z5. JWST+HST photometry is shown via data points with error bars (black: S/N$>$2; light gray: S/N$<$2), which are best fitted by galaxy templates in blue. The uncertainty of the best-fit template is shown in light blue, and the best template flux in filters is shown with filled circles.
On the right side of each panel, we show PDF(z) from EAZY with the best-fitted $z_{\rm phot}$ labeled.
\label{sed}
}
\end{figure*}

\section{Data, measurements, and selection}

\subsection{JWST+HST data processing and measurements}

We use the reduced and calibrated JWST+HST image products, and photometric and photo-$z$ catalogs that are publicly available\footnote{\href{https://s3.amazonaws.com/grizli-v2/JwstMosaics/v4/index.html}{https://s3.amazonaws.com/grizli-v2/JwstMosaics/v4/index.html}}. 
The JWST data reduction and calibration follow the same pipeline applied for data in \cite{Naidu2022z10}, \cite{Bradley2022z10}, \cite{Barrufet2022jwst}, \cite{Fujimoto2022z16}, and \cite{Bezanson2022UNCOVER}.
A detailed description of the pipeline will be provided in Brammer et al. (in prep.). Briefly, we retrieved the pipeline-calibrated level-2 NIRCam products from the Mikulski Archive for Space Telescopes (MAST), then further calibrated the data and processed them as mosaics using the \textsc{Grizli} package \citep{Brammer2021Grizli}. The \textsc{Grizli} pipeline masks various artifacts and models wisps, calibrates for a flat-field, subtracts large-scale sky background, and aligns the fully calibrated images to stars from the Gaia DR3 catalog \citep{Gaia2022}. The NIRCam photometric zero-point correction was applied and the derived photometric zeropoints are consistent with those derived by other JWST Early Release Science (ERS) teams \citep{Boyer2022NIRCam,Nardiello2022NIRCam}. The NIRCam images of short wavelength channels were processed to a final pixel scale of 0.02$''$, and 0.04$''$ per pixel for long wavelength channels. 
Sources were first extracted in the longest wavelength image using Source Extraction and Photometry (SEP, \citealt{Barbary2016}), and photometry was measured within an aperture of $0.5''$ diameter on the position from the first extraction. The measurement was performed in all JWST images with aperture correction.
The HST images in these fields were also processed with \textsc{Grizli} and the photometry was measured in an identical way.
The photometric redshifts were computed by fitting all available JWST+HST photometry with EAZY \citep{Brammer2008EAZY}. 
The catalog in the Cosmic Evolution Early Release Science Survey (CEERS) field \citep{Finkelstein2022CEERS} consists of 14 band photometry of which seven are from JWST (F115W, F150W, F200W, F277W, F356W, F410W, and F444W), and the remaining seven are from HST (F105W, F125W, F140W, F160W, F435W, F606W, and F814W). 
Spectroscopic redshifts are available for 748 sources in the CEERS catalog, which were collected from literature surveys of DEEP2/3 \citep{Cooper2011DEEP3,Newman2013DEEP2}, \cite{Masters2017}, MOSDEF \citep{Shivaei2018MOSDEF}, and SDSS DR15 \citep{Aguado2019SDSSdr15}. A comparison of the photo$-z$ estimates to the spectroscopic redshifts yield negligible systematics and a dispersion of $\Delta z/(1+z)= 2.68\%$.

For validation, the candidates have been examined from an independent photometric extraction by Weaver et al. (in prep.) using the \textsc{Grizli} mosaics that are point spread function PSF-matched to F160W. Sources were identified from a noise-equalized stacking of F277W, F356W, and F444W using SEP. Photometry was extracted using SEP in 0.7" diameter circular apertures on all available HST and JWST bands. Following \citet{Whitaker2011}, noise properties were measured in $>1000$ empty apertures. Photometry and associated uncertainties were then corrected to the total flux based on kron radii measured on the F444W mosaic. The two versions of photometry are found to be consistent, and we continue to use the public catalogs from Brammer et al. in following analysis.

\subsection{Selection}
CGG-z5, a compact galaxy group centered at RA 214.812 and Dec 52.8368,
was selected as the highest galaxy overdensity at $z>2$ in the JWST catalogs produced by Brammer et al. Using \textsc{GalCluster} \citep{Sillassen2022}, we mapped the distance of the fifth nearest neighbor $\Sigma_{\rm 5th}$ in redshift bins at $z=2$--9 in JWST fields of 208~arcmin$^2$, including CEERS-EGS, GLASS, SMACS-0723, Sunrise arc, J1235,
MACS-0647, Windhorst NEP TDF, NEP 2MASS star, and Stephan's quintet. In order to filter out largely uncertain photometric redshifts, $z_{\rm phot}$, in each redshift bin, $z_{\rm bin}$, only sources with a 16th percentile of $z_{\rm phot}$ $z_{\rm 16th}>z_{\rm bin} - 8\%\times(1+z_{\rm bin})$ and $z_{\rm 84th}<z_{\rm bin} + 8\%\times(1+z_{\rm bin})$ were preselected for overdensity analysis -- with 8\% being the 3$\sigma$ uncertainty of $1+z$ from comparison to spec-z (see Sect. 2.1). CGG-z5 is found with the highest overdensity in $\Sigma_{\rm 5th}$ which is 6.6$\sigma$ larger than the mean galaxy density at $5<z<6$ in the CEERS field. As shown in Fig.~\ref{img}, it hosts a compact group of galaxies, where six candidate members with $5.0<z_{\rm phot}<5.4$ are assembled in a $1.5''\times3''$ area ($\sim$10$\times$20~kpc$^2$ at $z=5.2$). 
We note that the six sources were first identified in F444W and are not blended in the image (Fig.~\ref{img}). The shorter wavelength photometry was measured within a fixed aperture on the position of F444W sources, thus the photometry measurements are free of blending issues and reliable for this sample.
To double-check this, we adopted an alternative overdensity analysis algorithm from \cite{Brinch2022cluster} and reconfirmed the high galaxy overdensity in CGG-z5.

We note that EAZY is a nonparametric code that fits SEDs using a linear combination of 12 preselected flexible stellar population synthesis (FSPS) templates. It is efficient at deriving photo-$z$ for large samples, while further detailed modeling is required for deriving physical parameters of individual sources.
To this end, we ran \textsc{Bagpipes} \citep{Carnall2018Bagpipes} SED fitting with JWST+HST photometry. Following \cite{Heintz2022z8}, we assumed a constant star formation history (SFH) and the attenuation curve of \citet{Salim2018attenuation}. We also adopted radiation fields in the range of $-3<logU<-1$, a metallicity grid from 0 to solar metallicity ($Z_\odot$), an $A_{V}$ grid from 0 to 2, and an age grid from 1~Myr to 2~Gyr. 
We found excellent agreement in $z_{\rm phot}$ between \textsc{Bagpipes} and EAZY (Fig.~\ref{z_compare}). We thus reran \textsc{Bagpipes} adopting the photometric redshifts as derived by EAZY. The results are listed in Table~\ref{tab:1}. We note that the stellar masses from \textsc{Bagpipes} are $\sim\times$1.5 lower compared to those inferred with EAZY for which the difference is caused by the different sets of templates and algorithms. We adopted this 0.18\,dex difference between the two codes as a systematic uncertainty in our stellar masses.

\begin{table*}[ht]
{
\caption{Candidate members in CGG-z5.}
\label{tab:1}
\renewcommand\arraystretch{1.4}
\centering
\begin{tabular}{ccccccc}
\hline\hline
     Name   &  $z_{\rm phot}$ & log($M_{*}/M_\odot$) & ${\rm SFR}$ &  $A_V$ & $Z$ & Age\\
            &  &  &    [M$_\odot$~yr$^{-1}$] &  & [$Z_\odot$] & [Gyr]\\
 \hline
CGG-z5.a  & $5.24^{+0.03}_{-0.10}$    &   $9.81^{+0.04}_{-0.04}$  &  $52^{+8}_{-6}$  & $0.42^{+0.06}_{-0.05}$ & $0.93^{+0.04}_{-0.05}$  &  $0.16^{+0.04}_{-0.03}$ \\
CGG-z5.b  & $5.04^{+0.20}_{-0.04}$    &   $9.0^{+0.07}_{-0.05}$   &  $12^{+2}_{-2}$   & $0.48^{+0.06}_{-0.06}$ & $0.44^{+0.13}_{-0.12}$  &  $0.09^{+0.03}_{-0.02}$ \\
CGG-z5.c  & $5.06^{+0.20}_{-0.04}$    &   $8.36^{+0.35}_{-0.17}$   &  $2^{+1}_{-1}$   & $0.48^{+0.11}_{-0.12}$    & $0.32^{+0.30}_{-0.21}$  &  $0.14^{+0.27}_{-0.08}$\\
CGG-z5.d  & $5.22^{+0.03}_{-0.11}$    &   $8.71^{+0.09}_{-0.07}$   &  $6^{+1}_{-1}$   & $0.20^{+0.04}_{-0.04}$   & $0.73^{+0.12}_{-0.14}$  &  $0.07^{+0.02}_{-0.02}$ \\
CGG-z5.e  & $5.10^{+0.10}_{-0.07}$    &   $8.52^{+0.22}_{-0.13}$  &  $3^{+1}_{-1}$   &  $0.39^{+0.08}_{-0.10}$ &  $0.25^{+0.17}_{-0.09}$  &  $0.14^{+0.18}_{-0.06}$ \\
CGG-z5.f  & $5.30^{+0.08}_{-0.32}$    &  $9.20^{+0.11}_{-0.09}$ &  $8^{+2}_{-1}$   & $0.27^{+0.06}_{-0.06}$   &  $0.57^{+0.22}_{-0.20}$  &  $0.25^{+0.13}_{-0.09}$ \\
\hline
\end{tabular}\\
}
{Notes: The $z_{\rm phot}$ results are from EAZY SED fitting with JWST+HST photometry, and the remaining parameters were computed by \textsc{Bagpipes} at the EAZY $z_{\rm phot}$ assuming a constant SFH and a Salim attenuation curve. Errors are quoted from the 16th and 84th quartiles without including a systematic uncertainty of $\sim$0.18\,dex in stellar mass.
} 
\end{table*}

\begin{figure*}[ht]
\setlength{\abovecaptionskip}{-0.1cm}
\setlength{\belowcaptionskip}{-0.2cm}
\centering
\includegraphics[width=0.9\textwidth]{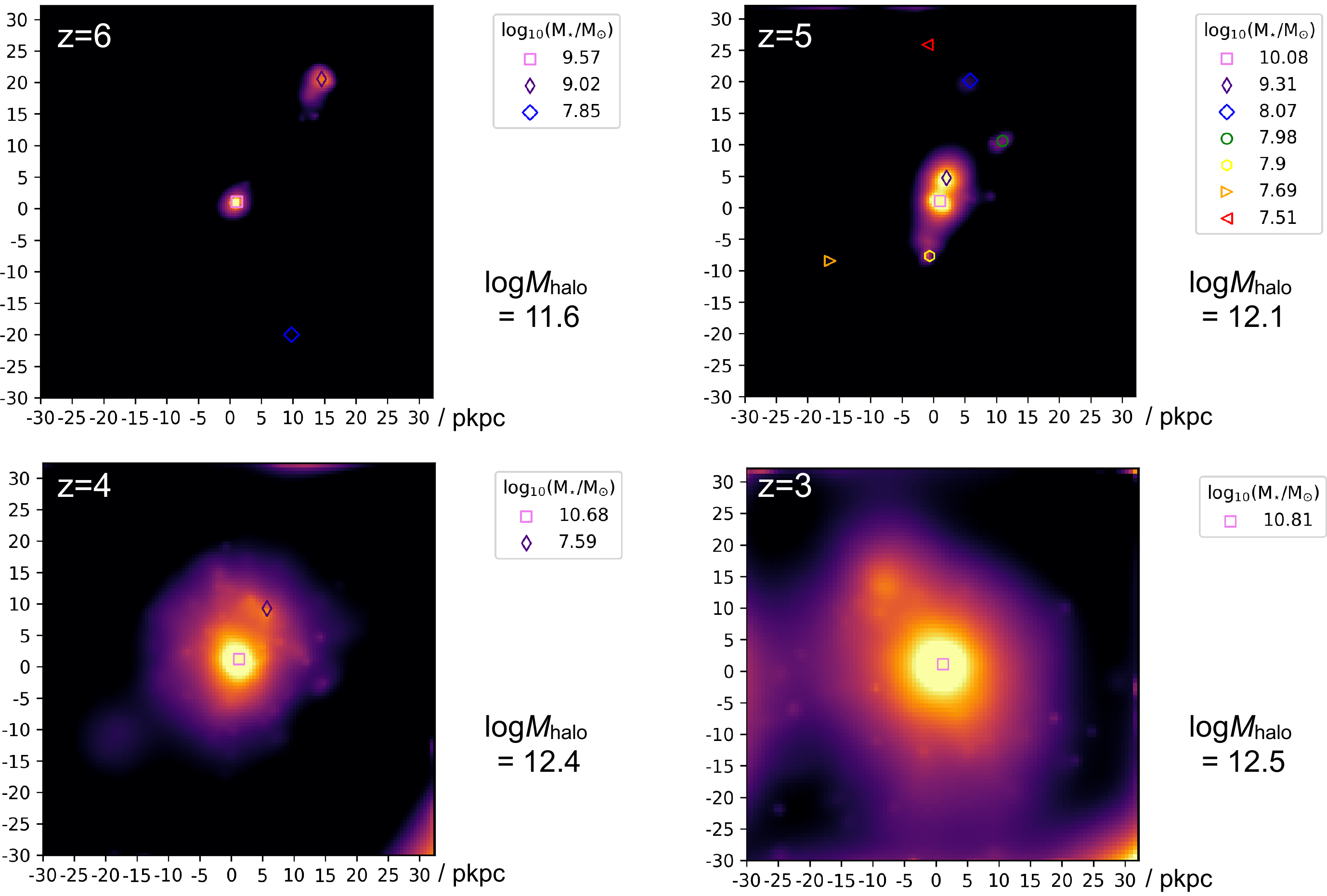}
\caption{%
Stellar mass map of a ``CGG-z5"-like group in the \textsc{Eagle} simulations. 
Stellar mass of individual galaxies are shown in the legend with halo mass $M_{\rm halo}=(M_{200}/M_{\odot})$.
The simulations suggest that CGG-z5 will form a single massive galaxy by $z=3$.
\label{simu}
}
\end{figure*}

\section{Results}

\subsection{A compact galaxy group at $z\sim5.2$}

We present the JWST NIRCam color images of CGG-z5 in Fig.~\ref{img}, and show HST and JWST images in Fig.~\ref{cutout}. The six candidate members all drop out in HST F435W and F606W data, they start to emerge in F814W, and are all well detected in NIRCam bands with similar colors. These properties are consistent with that of Lyman break galaxies (LBGs) at $z\sim5$. 

In Fig.~\ref{sed}, we show the best-fitted SEDs from EAZY. The 14 band photometry of JWST+HST are well fitted with a single peak at $z\sim5.2$ in the redshift probability distribution function PDF(z). 
All of the six galaxies agree well at $z\approx5.2$ within a $1\sigma$ uncertainty (Sect. 2.2).
We note that the excesses in F277W and F410M exhibited in the six SEDs support the $z\sim5.2$ solutions, as the F277W band brackets multiple strong lines [OII]+H$\gamma$+H$\beta$+[OIII] and F410M flux is boosted by strong H$\alpha$+[NII].
The best-fitted templates are all LBG-like with low dust attenuation, which is consistent with the nondetection in MIPS, Herschel, SCUBA2, and radio images. 
For a sanity check, we compared our results to two alternative versions of photo-z, one computed from an independent photometric extraction by Weaver et al. and the other one from \textsc{Bagpipes} (see Sect. 2.2). We found that the three versions of photo-z are in excellent agreement (Fig.~\ref{z_compare}).
These together suggest that CGG-z5 is a compact group of galaxies at $z\sim5.2$.

We list the results of EAZY and \textsc{Bagpipes} SED fitting results in Table~\ref{tab:1}. The most massive member CGG-z5.a has a stellar mass $M_*\sim10^{9.8}~M_\odot$ and a star formation rate (SFR) $\sim50~M_\odot$~yr$^{-1}$, placed $\sim2\times$ above,  yet still within the scatter of the main sequence at $z\sim5$ \citep{Schreiber2015}. The satellites around CGG-z5.a are less massive down to log$(M_*/M_\odot)\approx8.4$ with low SFRs $\approx 2-12~M_\odot$~yr$^{-1}$. 
The stellar mass ratios between the center and satellite galaxies are larger than 4:1, suggesting minor mergers.
Interestingly, CGG-z5.a is resolved into three components in the high resolution NIRCam images (Fig.~\ref{img}, left), and CGG-z5.d is resolved into two, suggesting ongoing mergers, star-forming clumps, or extra low-mass satellites.

CGG-z5 has some unique features with respect to previously identified compact structures. For example, compared to the compact group HPC1001 at $z=3.61$ (\citealt{Sillassen2022}, and in prep.), CGG-z5 has a higher projected number density by a factor of $\sim3$ and is less massive by an order of magnitude in stellar mass. 
AzTEC5, a starbursting group at $z\sim3.6$ hosting four candidate members \citep{Gomez-Guijarro2018}, has a total stellar mass and a SFR that are $3\times$ and $20\times$ higher compared to that of CGG-z5, respectively.
Another interesting case is the $z=2.94$ SDSS J165202.64+172852.3, which was recently observed by the NIRSpec IFU \citep{Wylezalek2022jwst}. This structure  hosts a red quasar and several companion galaxies in a scale of 10--15\,kpc. However, CGG-z5 is at a higher redshift, hosts hosts two times more galaxies (accounting for the resolved substructures of CGG-z5.a and d, see Fig.~\ref{img}, left), and the center galaxy is less massive than the quasar by 1.4--2.4 orders of magnitude.
Recently, a lensed protocluster at $z\sim7.9$ was  discovered with JWST in the Abell 2744 field \citep{Morishita2022z8cluster}, where the most overdense region contains six galaxies within a projected distance of 60\,kpc. It is at much higher redshift, but still less compact than CGG-z5 by a factor of 20 in the projected comoving number density.

\subsection{Detectability}

One could contemplate why this structure was previously undetected despite the breadth of the existing ancillary data. To answer this question, we examined CGG-z5 in the HST images and the public catalogs \citep{Skelton2014,Stefanon2017EGS}. As shown in Fig.~\ref{cutout}, the whole structure is totally invisible at wavelengths shorter than 0.8$\mu$m (indicative of the presence of a Lyman break), two members of the structure are marginally detected in F814W, and three  (CGG-z5.a, b, and d) are detected in the HST F160W image. These three sources are included in the 3D-HST catalog \citep{Skelton2014}, however with largely uncertain photo$-z$ estimates spanning from $z=$3.2--5.5. 
In the EGS-CANDELS catalog from \cite{Stefanon2017EGS}, four sources (CGG-z5.a, b, d, and f) are found in this group but again with largely dispersed $z_{\rm phot}=$1.2--5.0.
At the same time, while CGG-z5 is detected in the IRAC images, it is also severely hampered by blending due to its compactness with respect to the IRAC beam. 

In contrast, the state-of-the-art JWST imaging detects and resolves six members at seven wavelengths (1.15--4.44~$\mu$m), including three low-mass members previously undetected in HST images. JWST photometry significantly improved the constraints on the rest-frame UV slope and Balmer realm, while the nondetection in HST/F435W and F606W anchored the Lyman break. These combined remarkably reduced the photo$-z$ uncertainties. Moreover, the F277W and F410M bands perfectly bracket the features of strong emission lines at $z\sim5$, and the implicit flux excesses in the two bands further narrowed the redshift range. 
In Fig.~\ref{zPDF}, we compared the PDF(z) with and without JWST data, demonstrating that the identification of such faint structures, just one billion years after the Big Bang, can only be achieved with JWST.

\section{Discussion}

Admittedly, the nature  of such a compact and relatively low-mass structure at $z\sim5$ is unclear, with a range of possible evolutionary stages and paths: 
(1) a merging system where the low-mass satellites will merge into the central member of the structure, eventually forming  a single, massive galaxy; 
(2) a forming cluster core, which keeps growing via merger and cold gas accretion \citep{Dekel2013,Daddi2021Lya,Daddi2022Lya} and eventually forms a massive protocluster by $z=1$;
and (3) a filamentary structure of galaxies in the line of sight.
While spectroscopic follow-up observations (e.g., Keck/DEIMOS and JWST/NIRSpec) are instrumental to disentangle between these scenarios, we can resort to hydrodynamical simulations in the meantime.

In order to better understand the nature and evolution of CGG-z5, we searched for similar structures in the \textsc{Eagle} simulations \cite[]{Crain2015,Schaye2015}, a suite of smooth particle hydrodynamic (SPH) simulations. The free parameters of the model were calibrated to reproduce the stellar mass function, stellar mass - black hole mass relation, and the galaxy disk sizes at $z\sim0.1$. The model was also shown to reproduce many low-high redshift observational results which were not used in the calibration (e.g., \citealt{Furlong2015simu,Trayford2015,Katsianis2017EAGLE}). For this work we used the fiducial reference volume, a periodic box of 100 comoving megaparsecs per side.     
To find similar structures, we first searched for galaxies more massive than $M_*>10^{9.5}~M_{\odot}$ at $z\sim5$. From the sample, we then selected those with more than three nearby galaxies within a 30 proper kiloparsec radius. Only nearby galaxies with stellar mass greater than $10^{7}$ $M_{\odot}$ (all the galaxies have more than 100 stars and gas particles combined) were considered as companions. With this selection criteria, we found 14 such structures in the \textsc{Eagle} reference volume \footnote{The 3D stellar maps of the 14 \textsc{Eagle} structures are publicaly available here: \href{https://drive.google.com/drive/folders/18b4e7EIjbNgwOgYMLIjfgsIOQUrgua5m}
{https://drive.google.com/drive/folders/18b4e7EIjbNgwOgYMLIj\\fgsIOQUrgua5m}}. 
We traced their evolution with time by following the most bound star particles belonging to the different structures toward lower redshift, while the progenitors were traced using the most bound dark matter particles. As an example, in Fig.~\ref{simu} we show the one with closest resemblance to CGG-z5; it contains seven members and the two most massive members have stellar masses similar to those found in CGG-z5. 

As shown in Fig.~\ref{simu}, the simulated structure has a halo mass of log$(M_{\rm halo}/M_\odot)=12.1$ at $z=5$, which is consistent with the estimate of $0.9\times10^{12}~M_\odot$ for CGG-z5 from the stellar-mass-to-halo-mass relation in \cite{Shuntov2022}. Intriguingly, most of these members merge into the central galaxy by $z=4$, and form a single galaxy with log$(M_*/M_\odot)\sim10.7$. 
The stellar mass will keep growing to log$(M_*/M_\odot)\sim10.8$ at $z=3$ and log$(M_*/M_\odot)\sim11.5$ at $z=1$.
The other 13 simulated structures share the same fate, merging into a single, central, massive galaxy by $z=4-3$ with log$(M_*/M_\odot)\gtrsim11$ at $z\sim1$. It is thus tempting to speculate that the CGG-z5 system will also follow a similar evolutionary path, and we captured the process of satellites merging into a single massive galaxy.

It is also interesting to notice that the merging timescale in the simulations is $\sim400~$Myrs, which is in line with the fact that similar compact and overdense structures (i.e., with  more than five members and overdensity $\Sigma_{\rm 5th}>5\sigma$) are not identified at lower $ z$ nor at higher $z$ in the JWST catalogs. Looking at Fig.~\ref{simu}, this can be explained if similar structures have merged into a single galaxy or group with fewer members by $z\sim4$ and if, at $z>5$, JWST observations are not deep enough the detect satellite galaxies of lower $M_{\ast}$, and therefore not meet our adopted $\Sigma_{\rm 5th}$ overdensity criteria. This again supports the idea that CGG-z5 is in an early and short-lived phase of massive galaxy formation before coalescence, a so-called proto-massive galaxy.

We should stress that similar overdense structures, usually associated with submillimeter galaxies (SMGs) or qusars (e.g., AzTEC5, \citealt{Gomez-Guijarro2018}; SDSS J165202.64+172852.3, \citealt{Wylezalek2022jwst}), have been found at lower redshifts ($z\sim3-5$) hosting more massive galaxies (log($M_{\ast}$) $\ge$ 11) and with higher SFRs with respect to CGG-z5, however. According to the evolutionary picture described above, these structures at $z\sim5$ should host an even higher overdensity of galaxies compared to CGG-z5 and should be detectable with JWST. Given the low number density of these massive structures, it is not surprising that we fail to detect any of those in our volume limited observations. 
Indeed, based on the number density of SMGs in the AS2UDS sample 4~deg$^{-2}$ at $z>4$ \citep{Stach2019,Gullberg2016SW,Riechers2020z5SFR}, the expected number of $z>4$ SMGs in the volume covered by the JWST observations in this study is  $<1$.
The upcoming COSMOS-Web JWST survey, covering 0.54~deg$^2$ \citep{Casey2022COSMOS-Web}, is a expected to capture such structures.

An interesting question that arises is whether the merging groups, such as CGG-z5, are the main avenue for the formation of massive galaxies at $z\sim2-3$. We can attempt to gain some initial insights by comparing the space density of overdense structures at $z\sim5$, with the number density of massive galaxies at $z\sim3$. Adopting a redshift bin of $4.5<z<5.5$, we find that the space density of galaxies with log$(M_*/M_\odot)>9.8$ and $\geq3$ companions within a radius of 30~kpc is $8\times10^{-2}$~cMpc$^{-3}$.  On the other hand, in our JWST catalogs, we have identified twenty galaxies at $2.9<z<3.1$  with log$(M_*/M_\odot)>10.8$, corresponding to a space density of 0.12~cMpc$^{-3}$. While hampered by uncertainties (e.g., cosmic variance) and selection biases, the rough agreement between the two estimates makes merging groups appealing as the main progenitors of massive galaxies at cosmic noon.

\begin{acknowledgements}
The Cosmic Dawn Center (DAWN) is funded by the Danish National Research Foundation under grant No. 140.
SJ acknowledges the financial support from the European Union's Horizon Europe research and innovation program under the Marie Sk\l{}odowska-Curie grant agreement No. 101060888.
GEM acknowledges financial support from the Villum Young Investigator grant 37440 and 13160. APV and TRG acknowledges support from the Carlsberg Foundation (grant no CF20-0534). 
This work used the DiRAC@Durham facility managed by the Institute for Computational Cosmology on behalf of the STFC DiRAC HPC Facility (www.dirac.ac.uk). 
The equipment was funded by BEIS capital funding via STFC capital grants ST/K00042X/1, ST/P002293/1, ST/R002371/1 and ST/S002502/1, Durham University and STFC operations grant ST/R000832/1. 
DiRAC is part of the National e-Infrastructure. The \textsc{Eagle} simulations were performed using the DiRAC-2 facility at Durham, managed by the ICC, and the PRACE facility Curie based in France at TGCC, CEA, Bruyeres-le-Chatel. 
We also wish to acknowledge the following open source software packages used in the analysis: {Astropy \citep{astropy:2013, astropy:2018, astropy:2022}}, {Matplotlib \citep{Hunter:2007}}, {Numpy \citep{Harris2020_numpy}}, {Scipy \citep{2020SciPy-NMeth}}, TOPCAT \citep{Taylor2005Topcat}. 
\end{acknowledgements}

\bibliographystyle{aa}
\bibliography{biblio}

\begin{thebibliography}{64}
\expandafter\ifx\csname natexlab\endcsname\relax\def\natexlab#1{#1}\fi

\bibitem[{{Aguado} {et~al.}(2019){Aguado}, {Ahumada}, {Almeida}, {Anderson},
  {Andrews}, {Anguiano}, {Aquino Ort{\'\i}z}, {Arag{\'o}n-Salamanca},
  {Argudo-Fern{\'a}ndez}, {Aubert}, {Avila-Reese}, {Badenes}, {Barboza
  Rembold}, {Barger}, {Barrera-Ballesteros}, {Bates}, {Bautista}, {Beaton},
  {Beers}, {Belfiore}, {Bernardi}, {Bershady}, {Beutler}, {Bird}, {Bizyaev},
  {Blanc}, {Blanton}, {Blomqvist}, {Bolton}, {Boquien}, {Borissova}, {Bovy},
  {Brandt}, {Brinkmann}, {Brownstein}, {Bundy}, {Burgasser}, {Byler}, {Cano
  Diaz}, {Cappellari}, {Carrera}, {Cervantes Sodi}, {Chen}, {Cherinka}, {Choi},
  {Chung}, {Coffey}, {Comerford}, {Comparat}, {Covey}, {da Silva Ilha}, {da
  Costa}, {Dai}, {Damke}, {Darling}, {Davies}, {Dawson}, {de Sainte Agathe},
  {Deconto Machado}, {Del Moro}, {De Lee}, {Diamond-Stanic}, {Dom{\'\i}nguez
  S{\'a}nchez}, {Donor}, {Drory}, {du Mas des Bourboux}, {Duckworth}, {Dwelly},
  {Ebelke}, {Emsellem}, {Escoffier}, {Fern{\'a}ndez-Trincado}, {Feuillet},
  {Fischer}, {Fleming}, {Fraser-McKelvie}, {Freischlad}, {Frinchaboy}, {Fu},
  {Galbany}, {Garcia-Dias}, {Garc{\'\i}a-Hern{\'a}ndez}, {Garma Oehmichen},
  {Geimba Maia}, {Gil-Mar{\'\i}n}, {Grabowski}, {Gu}, {Guo}, {Ha},
  {Harrington}, {Hasselquist}, {Hayes}, {Hearty}, {Hernandez Toledo}, {Hicks},
  {Hogg}, {Holley-Bockelmann}, {Holtzman}, {Hsieh}, {Hunt}, {Hwang},
  {Ibarra-Medel}, {Jimenez Angel}, {Johnson}, {Jones}, {J{\"o}nsson},
  {Kinemuchi}, {Kollmeier}, {Krawczyk}, {Kreckel}, {Kruk}, {Lacerna}, {Lan},
  {Lane}, {Law}, {Lee}, {Li}, {Lian}, {Lin}, {Lin}, {Lintott}, {Long},
  {Longa-Pe{\~n}a}, {Mackereth}, {de la Macorra}, {Majewski}, {Malanushenko},
  {Manchado}, {Maraston}, {Mariappan}, {Marinelli}, {Marques-Chaves},
  {Masseron}, {Masters}, {McDermid}, {Medina Pe{\~n}a}, {Meneses-Goytia},
  {Merloni}, {Merrifield}, {Meszaros}, {Minniti}, {Minsley}, {Muna}, {Myers},
  {Nair}, {Correa do Nascimento}, {Newman}, {Nitschelm}, {Olmstead}, {Oravetz},
  {Oravetz}, {Ortega Minakata}, {Pace}, {Padilla}, {Palicio}, {Pan}, {Pan},
  {Parikh}, {Parker}, {Peirani}, {Penny}, {Percival}, {Perez-Fournon},
  {Peterken}, {Pinsonneault}, {Prakash}, {Raddick}, {Raichoor}, {Riffel},
  {Riffel}, {Rix}, {Robin}, {Roman-Lopes}, {Rose}, {Ross}, {Rossi}, {Rowlands},
  {Rubin}, {S{\'a}nchez}, {S{\'a}nchez-Gallego}, {Sayres}, {Schaefer},
  {Schiavon}, {Schimoia}, {Schlafly}, {Schlegel}, {Schneider}, {Schultheis},
  {Seo}, {Shamsi}, {Shao}, {Shen}, {Shetty}, {Simonian}, {Smethurst}, {Sobeck},
  {Souter}, {Spindler}, {Stark}, {Stassun}, {Steinmetz}, {Storchi-Bergmann},
  {Stringfellow}, {Su{\'a}rez}, {Sun}, {Taghizadeh-Popp}, {Talbot}, {Tayar},
  {Thakar}, {Thomas}, {Tissera}, {Tojeiro}, {Troup}, {Unda-Sanzana},
  {Valenzuela}, {Vargas-Maga{\~n}a}, {V{\'a}zquez-Mata}, {Wake}, {Weaver},
  {Weijmans}, {Westfall}, {Wild}, {Wilson}, {Woods}, {Yan}, {Yang}, {Zamora},
  {Zasowski}, {Zhang}, {Zheng}, {Zheng}, {Zhu}, {Zinn}, \&
  {Zou}}]{Aguado2019SDSSdr15}
{Aguado}, D.~S., {Ahumada}, R., {Almeida}, A., {et~al.} 2019, \apjs, 240, 23

\bibitem[{{Astropy Collaboration} {et~al.}(2022){Astropy Collaboration},
  {Price-Whelan}, {Lim}, {Earl}, {Starkman}, {Bradley}, {Shupe}, {Patil},
  {Corrales}, {Brasseur}, {N{"o}the}, {Donath}, {Tollerud}, {Morris},
  {Ginsburg}, {Vaher}, {Weaver}, {Tocknell}, {Jamieson}, {van Kerkwijk},
  {Robitaille}, {Merry}, {Bachetti}, {G{"u}nther}, {Aldcroft},
  {Alvarado-Montes}, {Archibald}, {B{'o}di}, {Bapat}, {Barentsen}, {Baz{'a}n},
  {Biswas}, {Boquien}, {Burke}, {Cara}, {Cara}, {Conroy}, {Conseil}, {Craig},
  {Cross}, {Cruz}, {D'Eugenio}, {Dencheva}, {Devillepoix}, {Dietrich},
  {Eigenbrot}, {Erben}, {Ferreira}, {Foreman-Mackey}, {Fox}, {Freij}, {Garg},
  {Geda}, {Glattly}, {Gondhalekar}, {Gordon}, {Grant}, {Greenfield}, {Groener},
  {Guest}, {Gurovich}, {Handberg}, {Hart}, {Hatfield-Dodds}, {Homeier},
  {Hosseinzadeh}, {Jenness}, {Jones}, {Joseph}, {Kalmbach}, {Karamehmetoglu},
  {Ka{l}uszy{'n}ski}, {Kelley}, {Kern}, {Kerzendorf}, {Koch}, {Kulumani},
  {Lee}, {Ly}, {Ma}, {MacBride}, {Maljaars}, {Muna}, {Murphy}, {Norman},
  {O'Steen}, {Oman}, {Pacifici}, {Pascual}, {Pascual-Granado}, {Patil},
  {Perren}, {Pickering}, {Rastogi}, {Roulston}, {Ryan}, {Rykoff}, {Sabater},
  {Sakurikar}, {Salgado}, {Sanghi}, {Saunders}, {Savchenko}, {Schwardt},
  {Seifert-Eckert}, {Shih}, {Jain}, {Shukla}, {Sick}, {Simpson},
  {Singanamalla}, {Singer}, {Singhal}, {Sinha}, {Sip{H{o}}cz}, {Spitler},
  {Stansby}, {Streicher}, {{S}umak}, {Swinbank}, {Taranu}, {Tewary},
  {Tremblay}, {Val-Borro}, {Van Kooten}, {Vasovi{'c}}, {Verma}, {de Miranda
  Cardoso}, {Williams}, {Wilson}, {Winkel}, {Wood-Vasey}, {Xue}, {Yoachim},
  {Zhang}, {Zonca}, \& {Astropy Project Contributors}}]{astropy:2022}
{Astropy Collaboration}, {Price-Whelan}, A.~M., {Lim}, P.~L., {et~al.} 2022,
  apj, 935, 167

\bibitem[{{Astropy Collaboration} {et~al.}(2018){Astropy Collaboration},
  {Price-Whelan}, {Sip{\H{o}}cz}, {G{\"u}nther}, {Lim}, {Crawford}, {Conseil},
  {Shupe}, {Craig}, {Dencheva}, {Ginsburg}, {Vand erPlas}, {Bradley},
  {P{\'e}rez-Su{\'a}rez}, {de Val-Borro}, {Aldcroft}, {Cruz}, {Robitaille},
  {Tollerud}, {Ardelean}, {Babej}, {Bach}, {Bachetti}, {Bakanov}, {Bamford},
  {Barentsen}, {Barmby}, {Baumbach}, {Berry}, {Biscani}, {Boquien}, {Bostroem},
  {Bouma}, {Brammer}, {Bray}, {Breytenbach}, {Buddelmeijer}, {Burke},
  {Calderone}, {Cano Rodr{\'\i}guez}, {Cara}, {Cardoso}, {Cheedella}, {Copin},
  {Corrales}, {Crichton}, {D'Avella}, {Deil}, {Depagne}, {Dietrich}, {Donath},
  {Droettboom}, {Earl}, {Erben}, {Fabbro}, {Ferreira}, {Finethy}, {Fox},
  {Garrison}, {Gibbons}, {Goldstein}, {Gommers}, {Greco}, {Greenfield},
  {Groener}, {Grollier}, {Hagen}, {Hirst}, {Homeier}, {Horton}, {Hosseinzadeh},
  {Hu}, {Hunkeler}, {Ivezi{\'c}}, {Jain}, {Jenness}, {Kanarek}, {Kendrew},
  {Kern}, {Kerzendorf}, {Khvalko}, {King}, {Kirkby}, {Kulkarni}, {Kumar},
  {Lee}, {Lenz}, {Littlefair}, {Ma}, {Macleod}, {Mastropietro}, {McCully},
  {Montagnac}, {Morris}, {Mueller}, {Mumford}, {Muna}, {Murphy}, {Nelson},
  {Nguyen}, {Ninan}, {N{\"o}the}, {Ogaz}, {Oh}, {Parejko}, {Parley}, {Pascual},
  {Patil}, {Patil}, {Plunkett}, {Prochaska}, {Rastogi}, {Reddy Janga},
  {Sabater}, {Sakurikar}, {Seifert}, {Sherbert}, {Sherwood-Taylor}, {Shih},
  {Sick}, {Silbiger}, {Singanamalla}, {Singer}, {Sladen}, {Sooley},
  {Sornarajah}, {Streicher}, {Teuben}, {Thomas}, {Tremblay}, {Turner},
  {Terr{\'o}n}, {van Kerkwijk}, {de la Vega}, {Watkins}, {Weaver}, {Whitmore},
  {Woillez}, {Zabalza}, \& {Astropy Contributors}}]{astropy:2018}
{Astropy Collaboration}, {Price-Whelan}, A.~M., {Sip{\H{o}}cz}, B.~M., {et~al.}
  2018, \aj, 156, 123

\bibitem[{{Astropy Collaboration} {et~al.}(2013){Astropy Collaboration},
  {Robitaille}, {Tollerud}, {Greenfield}, {Droettboom}, {Bray}, {Aldcroft},
  {Davis}, {Ginsburg}, {Price-Whelan}, {Kerzendorf}, {Conley}, {Crighton},
  {Barbary}, {Muna}, {Ferguson}, {Grollier}, {Parikh}, {Nair}, {Unther},
  {Deil}, {Woillez}, {Conseil}, {Kramer}, {Turner}, {Singer}, {Fox}, {Weaver},
  {Zabalza}, {Edwards}, {Azalee Bostroem}, {Burke}, {Casey}, {Crawford},
  {Dencheva}, {Ely}, {Jenness}, {Labrie}, {Lim}, {Pierfederici}, {Pontzen},
  {Ptak}, {Refsdal}, {Servillat}, \& {Streicher}}]{astropy:2013}
{Astropy Collaboration}, {Robitaille}, T.~P., {Tollerud}, E.~J., {et~al.} 2013,
  \aap, 558, A33

\bibitem[{{Barbary}(2016)}]{Barbary2016}
{Barbary}, K. 2016, {Extinction V0.3.0}, Zenodo

\bibitem[{{Barrufet} {et~al.}(2022){Barrufet}, {Oesch}, {Weibel}, {Brammer},
  {Bezanson}, {Bouwens}, {Fudamoto}, {Gonzalez}, {Illingworth}, {Heintz},
  {Holden}, {Labbe}, {Magee}, {Naidu}, {Nelson}, {Stefanon}, {Smit}, {van
  Dokkum}, {Weaver}, \& {Williams}}]{Barrufet2022jwst}
{Barrufet}, L., {Oesch}, P.~A., {Weibel}, A., {et~al.} 2022, arXiv e-prints,
  arXiv:2207.14733

\bibitem[{{Benson} {et~al.}(2012){Benson}, {Borgani}, {De Lucia},
  {Boylan-Kolchin}, \& {Monaco}}]{Benson2012MergerTree}
{Benson}, A.~J., {Borgani}, S., {De Lucia}, G., {Boylan-Kolchin}, M., \&
  {Monaco}, P. 2012, \mnras, 419, 3590

\bibitem[{{Bezanson} {et~al.}(2022){Bezanson}, {Labbe}, {Whitaker}, {Leja},
  {Price}, {Franx}, {Brammer}, {Marchesini}, {Zitrin}, {Wang}, {Weaver},
  {Furtak}, {Atek}, {Coe}, {Cutler}, {Dayal}, {van Dokkum}, {Feldmann},
  {Forster Schreiber}, {Fujimoto}, {Geha}, {Glazebrook}, {de Graaff}, {Greene},
  {Juneau}, {Kassin}, {Kriek}, {Khullar}, {Maseda}, {Mowla}, {Muzzin},
  {Nanayakkara}, {Nelson}, {Oesch}, {Pacifici}, {Pan}, {Papovich}, {Setton},
  {Shapley}, {Smit}, {Stefanon}, {Taylor}, \& {Williams}}]{Bezanson2022UNCOVER}
{Bezanson}, R., {Labbe}, I., {Whitaker}, K.~E., {et~al.} 2022, arXiv e-prints,
  arXiv:2212.04026

\bibitem[{{Boyer} {et~al.}(2022){Boyer}, {Anderson}, {Gennaro}, {Geha},
  {Wingfield McQuinn}, {Tollerud}, {Correnti}, {Brenner Newman}, {Cohen},
  {Kallivayalil}, {Beaton}, {Cole}, {Dolphin}, {Kalirai}, {Sandstrom},
  {Savino}, {Skillman}, {Weisz}, \& {Williams}}]{Boyer2022NIRCam}
{Boyer}, M.~L., {Anderson}, J., {Gennaro}, M., {et~al.} 2022, Research Notes of
  the American Astronomical Society, 6, 191

\bibitem[{{Boylan-Kolchin} {et~al.}(2009){Boylan-Kolchin}, {Springel}, {White},
  {Jenkins}, \& {Lemson}}]{Boylan-Kolchin2009}
{Boylan-Kolchin}, M., {Springel}, V., {White}, S. D.~M., {Jenkins}, A., \&
  {Lemson}, G. 2009, \mnras, 398, 1150

\bibitem[{{Bradley} {et~al.}(2022){Bradley}, {Coe}, {Brammer}, {Furtak},
  {Larson}, {Andrade-Santos}, {Bhatawdekar}, {Bradac}, {Broadhurst}, {Carnall},
  {Conselice}, {Diego}, {Frye}, {Fujimoto}, {Y. -Y Hsiao}, {Hutchison}, {Jung},
  {Mahler}, {McCandliss}, {Oguri}, {Postman}, {Sharon}, {Trenti}, {Vanzella},
  {Welch}, {Windhorst}, \& {Zitrin}}]{Bradley2022z10}
{Bradley}, L.~D., {Coe}, D., {Brammer}, G., {et~al.} 2022, arXiv e-prints,
  arXiv:2210.01777

\bibitem[{{Brammer} \& {Matharu}(2021)}]{Brammer2021Grizli}
{Brammer}, G. \& {Matharu}, J. 2021, {gbrammer/grizli: Release 2021}, Zenodo

\bibitem[{{Brammer} {et~al.}(2008){Brammer}, {van Dokkum}, \&
  {Coppi}}]{Brammer2008EAZY}
{Brammer}, G.~B., {van Dokkum}, P.~G., \& {Coppi}, P. 2008, \apj, 686, 1503

\bibitem[{{Brinch} {et~al.}(2022){Brinch}, {Greve}, {Weaver}, {Brammer},
  {Ilbert}, {Shuntov}, {Jin}, {Liu}, {Gim{\'e}nez-Arteaga}, {Casey},
  {Davidson}, {Fujimoto}, {Koekemoer}, {Kokorev}, {Magdis}, {McCracken},
  {McPartland}, {Mobasher}, {Sanders}, {Toft}, {Valentino}, {Zamorani}, \&
  {Zavala}}]{Brinch2022cluster}
{Brinch}, M., {Greve}, T.~R., {Weaver}, J.~R., {et~al.} 2022, arXiv e-prints,
  arXiv:2210.17334

\bibitem[{{Carnall} {et~al.}(2018){Carnall}, {McLure}, {Dunlop}, \&
  {Dav{\'e}}}]{Carnall2018Bagpipes}
{Carnall}, A.~C., {McLure}, R.~J., {Dunlop}, J.~S., \& {Dav{\'e}}, R. 2018,
  \mnras, 480, 4379

\bibitem[{{Casey} {et~al.}(2022){Casey}, {Kartaltepe}, {Drakos}, {Franco},
  {Ilbert}, {Rose}, {Cox}, {Nightingale}, {Robertson}, {Silverman},
  {Koekemoer}, {Massey}, {McCracken}, {Rhodes}, {Akins}, {Amvrosiadis},
  {Arango-Toro}, {Bagley}, {Capak}, {Champagne}, {Chartab}, {Chavez Ortiz},
  {Cooke}, {Cooper}, {Darvish}, {Ding}, {Faisst}, {Finkelstein}, {Fujimoto},
  {Gentile}, {Gillman}, {Gould}, {Gozaliasl}, {Harish}, {Hayward}, {He},
  {Hemmati}, {Hirschmann}, {Jin}, {Khostovan}, {Kokorev}, {Lambrides},
  {Laigle}, {Leung}, {Liu}, {Liaudat}, {Long}, {Magdis}, {Mahler}, {Mainieri},
  {Manning}, {Maraston}, {Martin}, {McCleary}, {McKinney}, {McPartland},
  {Mobasher}, {Pattnaik}, {Renzini}, {Rich}, {Sanders}, {Sattari},
  {Scognamiglio}, {Scoville}, {Sheth}, {Shuntov}, {Sparre}, {Suzuki}, {Talia},
  {Toft}, {Trakhtenbrot}, {Urry}, {Valentino}, {Vanderhoof}, {Vardoulaki},
  {Weaver}, {Whitaker}, {Wilkins}, {Yang}, \& {Zavala}}]{Casey2022COSMOS-Web}
{Casey}, C.~M., {Kartaltepe}, J.~S., {Drakos}, N.~E., {et~al.} 2022, arXiv
  e-prints, arXiv:2211.07865

\bibitem[{{Chabrier}(2003)}]{Chabrier2003}
{Chabrier}, G. 2003, \pasp, 115, 763

\bibitem[{{Cooper} {et~al.}(2011){Cooper}, {Aird}, {Coil}, {Davis}, {Faber},
  {Juneau}, {Lotz}, {Nandra}, {Newman}, {Willmer}, \& {Yan}}]{Cooper2011DEEP3}
{Cooper}, M.~C., {Aird}, J.~A., {Coil}, A.~L., {et~al.} 2011, \apjs, 193, 14

\bibitem[{{Crain} {et~al.}(2015){Crain}, {Schaye}, {Bower}, {Furlong},
  {Schaller}, {Theuns}, {Dalla Vecchia}, {Frenk}, {McCarthy}, {Helly},
  {Jenkins}, {Rosas-Guevara}, {White}, \& {Trayford}}]{Crain2015}
{Crain}, R.~A., {Schaye}, J., {Bower}, R.~G., {et~al.} 2015, \mnras, 450, 1937

\bibitem[{{Daddi} {et~al.}(2022){Daddi}, {Rich}, {Valentino}, {Jin},
  {Delvecchio}, {Liu}, {Strazzullo}, {Neill}, {Gobat}, {Finoguenov},
  {Bournaud}, {Elbaz}, {Kalita}, {O'Sullivan}, \& {Wang}}]{Daddi2022Lya}
{Daddi}, E., {Rich}, R.~M., {Valentino}, F., {et~al.} 2022, \apjl, 926, L21

\bibitem[{{Daddi} {et~al.}(2021){Daddi}, {Valentino}, {Rich}, {Neill},
  {Gronke}, {O'Sullivan}, {Elbaz}, {Bournaud}, {Finoguenov}, {Marchal},
  {Delvecchio}, {Jin}, {Liu}, {Strazzullo}, {Calabro}, {Coogan}, {D'Eugenio},
  {Gobat}, {Kalita}, {Laursen}, {Martin}, {Puglisi}, {Schinnerer}, \&
  {Wang}}]{Daddi2021Lya}
{Daddi}, E., {Valentino}, F., {Rich}, R.~M., {et~al.} 2021, \aap, 649, A78

\bibitem[{{Dekel} {et~al.}(2013){Dekel}, {Zolotov}, {Tweed}, {Cacciato},
  {Ceverino}, \& {Primack}}]{Dekel2013}
{Dekel}, A., {Zolotov}, A., {Tweed}, D., {et~al.} 2013, \mnras, 435, 999

\bibitem[{{Finkelstein} {et~al.}(2022){Finkelstein}, {Bagley}, {Ferguson},
  {Wilkins}, {Kartaltepe}, {Papovich}, {Yung}, {Arrabal Haro}, {Behroozi},
  {Dickinson}, {Kocevski}, {Koekemoer}, {Larson}, {Le Bail}, {Morales},
  {Perez-Gonzalez}, {Burgarella}, {Dave}, {Hirschmann}, {Somerville}, {Wuyts},
  {Bromm}, {Casey}, {Fontana}, {Fujimoto}, {Gardner}, {Giavalisco}, {Grazian},
  {Grogin}, {Hathi}, {Hutchison}, {Jha}, {Jogee}, {Kewley}, {Kirkpatrick},
  {Long}, {Lotz}, {Pentericci}, {Pierel}, {Pirzkal}, {Ravindranath}, {Ryan},
  {Trump}, {Yang}, {Bhatawdekar}, {Bisigello}, {Buat}, {Calabro}, {Castellano},
  {Cleri}, {Cooper}, {Croton}, {Daddi}, {Dekel}, {Elbaz}, {Franco}, {Gawiser},
  {Holwerda}, {Huertas-Company}, {Jaskot}, {Leung}, {Lucas}, {Mobasher},
  {Pandya}, {Tacchella}, {Weiner}, \& {Zavala}}]{Finkelstein2022CEERS}
{Finkelstein}, S.~L., {Bagley}, M.~B., {Ferguson}, H.~C., {et~al.} 2022, arXiv
  e-prints, arXiv:2211.05792

\bibitem[{{Fujimoto} {et~al.}(2022){Fujimoto}, {Finkelstein}, {Burgarella},
  {Carilli}, {Buat}, {Casey}, {Ciesla}, {Tacchella}, {Zavala}, {Brammer},
  {Fudamoto}, {Ouchi}, {Valentino}, {Cooper}, {Dickinson}, {Franco},
  {Giavalisco}, {Hutchison}, {Kartaltepe}, {Koekemoer}, {Kojima}, {Larson},
  {Murphy}, {Papovich}, {P{\'e}rez-Gonz{\'a}lez}, {Somerville}, {Yoon},
  {Wilkins}, {Yung}, {Akins}, {Amor{\'\i}n}, {Arrabal Haro}, {Bagley},
  {Chworowsky}, {Cooper}, {Costantin}, {Daddi}, {Ferguson}, {Grogin},
  {Jim{\'e}nez-Andrade}, {Juneau}, {Kirkpatrick}, {Kocevski}, {Le Bail},
  {Long}, {Lucas}, {Magnelli}, {McKinney}, {Rose}, {Seill{\'e}}, {Simons}, \&
  {Weiner}}]{Fujimoto2022z16}
{Fujimoto}, S., {Finkelstein}, S.~L., {Burgarella}, D., {et~al.} 2022, arXiv
  e-prints, arXiv:2211.03896

\bibitem[{{Furlong} {et~al.}(2015){Furlong}, {Bower}, {Theuns}, {Schaye},
  {Crain}, {Schaller}, {Dalla Vecchia}, {Frenk}, {McCarthy}, {Helly},
  {Jenkins}, \& {Rosas-Guevara}}]{Furlong2015simu}
{Furlong}, M., {Bower}, R.~G., {Theuns}, T., {et~al.} 2015, \mnras, 450, 4486

\bibitem[{{Gaia Collaboration} {et~al.}(2022){Gaia Collaboration}, {Vallenari},
  {Brown}, {Prusti}, {de Bruijne}, {Arenou}, {Babusiaux}, {Biermann},
  {Creevey}, {Ducourant}, {Evans}, {Eyer}, {Guerra}, {Hutton}, {Jordi},
  {Klioner}, {Lammers}, {Lindegren}, {Luri}, {Mignard}, {Panem}, {Pourbaix},
  {Randich}, {Sartoretti}, {Soubiran}, {Tanga}, {Walton}, {Bailer-Jones},
  {Bastian}, {Drimmel}, {Jansen}, {Katz}, {Lattanzi}, {van Leeuwen}, {Bakker},
  {Cacciari}, {Casta{\~n}eda}, {De Angeli}, {Fabricius}, {Fouesneau},
  {Fr{\'e}mat}, {Galluccio}, {Guerrier}, {Heiter}, {Masana}, {Messineo},
  {Mowlavi}, {Nicolas}, {Nienartowicz}, {Pailler}, {Panuzzo}, {Riclet}, {Roux},
  {Seabroke}, {Sordo{\o}rcit}, {Th{\'e}venin}, {Gracia-Abril}, {Portell},
  {Teyssier}, {Altmann}, {Andrae}, {Audard}, {Bellas-Velidis}, {Benson},
  {Berthier}, {Blomme}, {Burgess}, {Busonero}, {Busso}, {C{\'a}novas}, {Carry},
  {Cellino}, {Cheek}, {Clementini}, {Damerdji}, {Davidson}, {de Teodoro},
  {Nu{\~n}ez Campos}, {Delchambre}, {Dell'Oro}, {Esquej},
  {Fern{\'a}ndez-Hern{\'a}ndez}, {Fraile}, {Garabato}, {Garc{\'\i}a-Lario},
  {Gosset}, {Haigron}, {Halbwachs}, {Hambly}, {Harrison}, {Hern{\'a}ndez},
  {Hestroffer}, {Hodgkin}, {Holl}, {Jan{\ss}en}, {Jevardat de Fombelle},
  {Jordan}, {Krone-Martins}, {Lanzafame}, {L{\"o}ffler}, {Marchal}, {Marrese},
  {Moitinho}, {Muinonen}, {Osborne}, {Pancino}, {Pauwels}, {Recio-Blanco},
  {Reyl{\'e}}, {Riello}, {Rimoldini}, {Roegiers}, {Rybizki}, {Sarro}, {Siopis},
  {Smith}, {Sozzetti}, {Utrilla}, {van Leeuwen}, {Abbas}, {{\'A}brah{\'a}m},
  {Abreu Aramburu}, {Aerts}, {Aguado}, {Ajaj}, {Aldea-Montero}, {Altavilla},
  {{\'A}lvarez}, {Alves}, {Anders}, {Anderson}, {Anglada Varela}, {Antoja},
  {Baines}, {Baker}, {Balaguer-N{\'u}{\~n}ez}, {Balbinot}, {Balog}, {Barache},
  {Barbato}, {Barros}, {Barstow}, {Bartolom{\'e}}, {Bassilana}, {Bauchet},
  {Becciani}, {Bellazzini}, {Berihuete}, {Bernet}, {Bertone}, {Bianchi},
  {Binnenfeld}, {Blanco-Cuaresma}, {Blazere}, {Boch}, {Bombrun}, {Bossini},
  {Bouquillon}, {Bragaglia}, {Bramante}, {Breedt}, {Bressan}, {Brouillet},
  {Brugaletta}, {Bucciarelli}, {Burlacu}, {Butkevich}, {Buzzi}, {Caffau},
  {Cancelliere}, {Cantat-Gaudin}, {Carballo}, {Carlucci}, {Carnerero},
  {Carrasco}, {Casamiquela}, {Castellani}, {Castro-Ginard}, {Chaoul},
  {Charlot}, {Chemin}, {Chiaramida}, {Chiavassa}, {Chornay}, {Comoretto},
  {Contursi}, {Cooper}, {Cornez}, {Cowell}, {Crifo}, {Cropper}, {Crosta},
  {Crowley}, {Dafonte}, {Dapergolas}, {David}, {David}, {de Laverny}, {De
  Luise}, {De March}, {De Ridder}, {de Souza}, {de Torres}, {del Peloso}, {del
  Pozo}, {Delbo}, {Delgado}, {Delisle}, {Demouchy}, {Dharmawardena}, {Di
  Matteo}, {Diakite}, {Diener}, {Distefano}, {Dolding}, {Edvardsson}, {Enke},
  {Fabre}, {Fabrizio}, {Faigler}, {Fedorets}, {Fernique}, {Fienga}, {Figueras},
  {Fournier}, {Fouron}, {Fragkoudi}, {Gai}, {Garcia-Gutierrez},
  {Garcia-Reinaldos}, {Garc{\'\i}a-Torres}, {Garofalo}, {Gavel}, {Gavras},
  {Gerlach}, {Geyer}, {Giacobbe}, {Gilmore}, {Girona}, {Giuffrida}, {Gomel},
  {Gomez}, {Gonz{\'a}lez-N{\'u}{\~n}ez}, {Gonz{\'a}lez-Santamar{\'\i}a},
  {Gonz{\'a}lez-Vidal}, {Granvik}, {Guillout}, {Guiraud},
  {Guti{\'e}rrez-S{\'a}nchez}, {Guy}, {Hatzidimitriou}, {Hauser}, {Haywood},
  {Helmer}, {Helmi}, {Sarmiento}, {Hidalgo}, {Hilger}, {H{\l}adczuk}, {Hobbs},
  {Holland}, {Huckle}, {Jardine}, {Jasniewicz}, {Jean-Antoine Piccolo},
  {Jim{\'e}nez-Arranz}, {Jorissen}, {Juaristi Campillo}, {Julbe}, {Karbevska},
  {Kervella}, {Khanna}, {Kontizas}, {Kordopatis}, {Korn}, {K{\'o}sp{\'a}l},
  {Kostrzewa-Rutkowska}, {Kruszy{\'n}ska}, {Kun}, {Laizeau}, {Lambert},
  {Lanza}, {Lasne}, {Le Campion}, {Lebreton}, {Lebzelter}, {Leccia}, {Leclerc},
  {Lecoeur-Taibi}, {Liao}, {Licata}, {Lindstr{\o}m}, {Lister}, {Livanou},
  {Lobel}, {Lorca}, {Loup}, {Madrero Pardo}, {Magdaleno Romeo}, {Managau},
  {Mann}, {Manteiga}, {Marchant}, {Marconi}, {Marcos}, {Marcos Santos},
  {Mar{\'\i}n Pina}, {Marinoni}, {Marocco}, {Marshall}, {Polo},
  {Mart{\'\i}n-Fleitas}, {Marton}, {Mary}, {Masip}, {Massari},
  {Mastrobuono-Battisti}, {Mazeh}, {McMillan}, {Messina}, {Michalik}, {Millar},
  {Mints}, {Molina}, {Molinaro}, {Moln{\'a}r}, {Monari}, {Mongui{\'o}},
  {Montegriffo}, {Montero}, {Mor}, {Mora}, {Morbidelli}, {Morel}, {Morris},
  {Muraveva}, {Murphy}, {Musella}, {Nagy}, {Noval}, {Oca{\~n}a}, {Ogden},
  {Ordenovic}, {Osinde}, {Pagani}, {Pagano}, {Palaversa}, {Palicio},
  {Pallas-Quintela}, {Panahi}, {Payne-Wardenaar}, {Pe{\~n}alosa Esteller},
  {Penttil{\"a}}, {Pichon}, {Piersimoni}, {Pineau}, {Plachy}, {Plum}, {Poggio},
  {Pr{\v{s}}a}, {Pulone}, {Racero}, {Ragaini}, {Rainer}, {Raiteri}, {Rambaux},
  {Ramos}, {Ramos-Lerate}, {Re Fiorentin}, {Regibo}, {Richards}, {Rios Diaz},
  {Ripepi}, {Riva}, {Rix}, {Rixon}, {Robichon}, {Robin}, {Robin}, {Roelens},
  {Rogues}, {Rohrbasser}, {Romero-G{\'o}mez}, {Rowell}, {Royer}, {Ruz Mieres},
  {Rybicki}, {Sadowski}, {S{\'a}ez N{\'u}{\~n}ez}, {Sagrist{\`a} Sell{\'e}s},
  {Sahlmann}, {Salguero}, {Samaras}, {Sanchez Gimenez}, {Sanna},
  {Santove{\~n}a}, {Sarasso}, {Schultheis}, {Sciacca}, {Segol}, {Segovia},
  {S{\'e}gransan}, {Semeux}, {Shahaf}, {Siddiqui}, {Siebert}, {Siltala},
  {Silvelo}, {Slezak}, {Slezak}, {Smart}, {Snaith}, {Solano}, {Solitro},
  {Souami}, {Souchay}, {Spagna}, {Spina}, {Spoto}, {Steele},
  {Steidelm{\"u}ller}, {Stephenson}, {S{\"u}veges}, {Surdej}, {Szabados},
  {Szegedi-Elek}, {Taris}, {Taylo}, {Teixeira}, {Tolomei}, {Tonello}, {Torra},
  {Torra}, {Torralba Elipe}, {Trabucchi}, {Tsounis}, {Turon}, {Ulla}, {Unger},
  {Vaillant}, {van Dillen}, {van Reeven}, {Vanel}, {Vecchiato}, {Viala},
  {Vicente}, {Voutsinas}, {Weiler}, {Wevers}, {Wyrzykowski}, {Yoldas}, {Yvard},
  {Zhao}, {Zorec}, {Zucker}, \& {Zwitter}}]{Gaia2022}
{Gaia Collaboration}, {Vallenari}, A., {Brown}, A.~G.~A., {et~al.} 2022, arXiv
  e-prints, arXiv:2208.00211

\bibitem[{{G{\'o}mez-Guijarro} {et~al.}(2018){G{\'o}mez-Guijarro}, {Toft},
  {Karim}, {Magnelli}, {Magdis}, {Jim{\'e}nez-Andrade}, {Capak}, {Fraternali},
  {Fujimoto}, {Riechers}, {Schinnerer}, {Smol{\v{c}}i{\'c}}, {Aravena},
  {Bertoldi}, {Cortzen}, {Hasinger}, {Hu}, {Jones}, {Koekemoer}, {Lee},
  {McCracken}, {Micha{\l}owski}, {Navarrete}, {Povi{\'c}}, {Puglisi},
  {Romano-D{\'\i}az}, {Sheth}, {Silverman}, {Staguhn}, {Steinhardt},
  {Stockmann}, {Tanaka}, {Valentino}, {van Kampen}, \&
  {Zirm}}]{Gomez-Guijarro2018}
{G{\'o}mez-Guijarro}, C., {Toft}, S., {Karim}, A., {et~al.} 2018, \apj, 856,
  121

\bibitem[{{Gullberg} {et~al.}(2016){Gullberg}, {Lehnert}, {De Breuck},
  {Branchu}, {Dannerbauer}, {Drouart}, {Emonts}, {Guillard}, {Hatch},
  {Nesvadba}, {Omont}, {Seymour}, \& {Vernet}}]{Gullberg2016SW}
{Gullberg}, B., {Lehnert}, M.~D., {De Breuck}, C., {et~al.} 2016, \aap, 591,
  A73

\bibitem[{{Harris} {et~al.}(2020){Harris}, {Millman}, {van der Walt},
  {Gommers}, {Virtanen}, {Cournapeau}, {Wieser}, {Taylor}, {Berg}, {Smith},
  {Kern}, {Picus}, {Hoyer}, {van Kerkwijk}, {Brett}, {Haldane}, {del R{\'\i}o},
  {Wiebe}, {Peterson}, {G{\'e}rard-Marchant}, {Sheppard}, {Reddy}, {Weckesser},
  {Abbasi}, {Gohlke}, \& {Oliphant}}]{Harris2020_numpy}
{Harris}, C.~R., {Millman}, K.~J., {van der Walt}, S.~J., {et~al.} 2020, \nat,
  585, 357

\bibitem[{{Heintz} {et~al.}(2022){Heintz}, {Gim{\'e}nez-Arteaga}, {Fujimoto},
  {Brammer}, {Espada}, {Gillman}, {Gonz{\'a}lez-L{\'o}pez}, {Greve},
  {Harikane}, {Hatsukade}, {Knudsen}, {Koekemoer}, {Kohno}, {Kokorev}, {Lee},
  {Magdis}, {Nelson}, {Rizzo}, {Sanders}, {Schaerer}, {Shapley}, {Strait},
  {Sun}, {Toft}, {Valentino}, {Vijayan}, {Watson}, {Bauer}, {Christiansen}, \&
  {Wilson}}]{Heintz2022z8}
{Heintz}, K.~E., {Gim{\'e}nez-Arteaga}, C., {Fujimoto}, S., {et~al.} 2022,
  arXiv e-prints, arXiv:2212.06877

\bibitem[{{Hirschmann} {et~al.}(2012){Hirschmann}, {Naab}, {Somerville},
  {Burkert}, \& {Oser}}]{Hirschmann2012simu}
{Hirschmann}, M., {Naab}, T., {Somerville}, R.~S., {Burkert}, A., \& {Oser}, L.
  2012, \mnras, 419, 3200

\bibitem[{{Hopkins} {et~al.}(2009){Hopkins}, {Somerville}, {Cox}, {Hernquist},
  {Jogee}, {Kere{\v{s}}}, {Ma}, {Robertson}, \& {Stewart}}]{Hopkins2009}
{Hopkins}, P.~F., {Somerville}, R.~S., {Cox}, T.~J., {et~al.} 2009, \mnras,
  397, 802

\bibitem[{Hunter(2007)}]{Hunter:2007}
Hunter, J.~D. 2007, Computing in Science \& Engineering, 9, 90

\bibitem[{{Katsianis} {et~al.}(2017){Katsianis}, {Blanc}, {Lagos}, {Tejos},
  {Bower}, {Alavi}, {Gonzalez}, {Theuns}, {Schaller}, \&
  {Lopez}}]{Katsianis2017EAGLE}
{Katsianis}, A., {Blanc}, G., {Lagos}, C.~P., {et~al.} 2017, \mnras, 472, 919

\bibitem[{{Klypin} {et~al.}(2011){Klypin}, {Trujillo-Gomez}, \&
  {Primack}}]{Klypin2011halo}
{Klypin}, A.~A., {Trujillo-Gomez}, S., \& {Primack}, J. 2011, \apj, 740, 102

\bibitem[{{Masters} {et~al.}(2017){Masters}, {Stern}, {Cohen}, {Capak},
  {Rhodes}, {Castander}, \& {Paltani}}]{Masters2017}
{Masters}, D.~C., {Stern}, D.~K., {Cohen}, J.~G., {et~al.} 2017, \apj, 841, 111

\bibitem[{{Morishita} {et~al.}(2022){Morishita}, {Roberts-Borsani}, {Treu},
  {Brammer}, {Mason}, {Trenti}, {Vulcani}, {Wang}, {Acebron}, {Bah{\'e}},
  {Bergamini}, {Boyett}, {Bradac}, {Calabr{\`o}}, {Castellano}, {Chen}, {De
  Lucia}, {Filippenko}, {Fontana}, {Glazebrook}, {Grillo}, {Henry}, {Jones},
  {Kelly}, {Koekemoer}, {Leethochawalit}, {Lu}, {Marchesini}, {Mascia},
  {Mercurio}, {Merlin}, {Metha}, {Nanayakkara}, {Nonino}, {Paris},
  {Pentericci}, {Santini}, {Strait}, {Vanzella}, {Windhorst}, {Rosati}, \&
  {Xie}}]{Morishita2022z8cluster}
{Morishita}, T., {Roberts-Borsani}, G., {Treu}, T., {et~al.} 2022, arXiv
  e-prints, arXiv:2211.09097

\bibitem[{{Naidu} {et~al.}(2022){Naidu}, {Oesch}, {van Dokkum}, {Nelson},
  {Suess}, {Brammer}, {Whitaker}, {Illingworth}, {Bouwens}, {Tacchella},
  {Matthee}, {Allen}, {Bezanson}, {Conroy}, {Labbe}, {Leja}, {Leonova},
  {Magee}, {Price}, {Setton}, {Strait}, {Stefanon}, {Toft}, {Weaver}, \&
  {Weibel}}]{Naidu2022z10}
{Naidu}, R.~P., {Oesch}, P.~A., {van Dokkum}, P., {et~al.} 2022, \apjl, 940,
  L14

\bibitem[{{Nardiello} {et~al.}(2022){Nardiello}, {Bedin}, {Burgasser},
  {Salaris}, {Cassisi}, {Griggio}, \& {Scalco}}]{Nardiello2022NIRCam}
{Nardiello}, D., {Bedin}, L.~R., {Burgasser}, A., {et~al.} 2022, \mnras, 517,
  484

\bibitem[{{Neistein} \& {Dekel}(2008)}]{Neistein_Dekel2008}
{Neistein}, E. \& {Dekel}, A. 2008, \mnras, 383, 615

\bibitem[{{Newman} {et~al.}(2013){Newman}, {Cooper}, {Davis}, {Faber}, {Coil},
  {Guhathakurta}, {Koo}, {Phillips}, {Conroy}, {Dutton}, {Finkbeiner}, {Gerke},
  {Rosario}, {Weiner}, {Willmer}, {Yan}, {Harker}, {Kassin}, {Konidaris},
  {Lai}, {Madgwick}, {Noeske}, {Wirth}, {Connolly}, {Kaiser}, {Kirby},
  {Lemaux}, {Lin}, {Lotz}, {Luppino}, {Marinoni}, {Matthews}, {Metevier}, \&
  {Schiavon}}]{Newman2013DEEP2}
{Newman}, J.~A., {Cooper}, M.~C., {Davis}, M., {et~al.} 2013, \apjs, 208, 5

\bibitem[{{Oser} {et~al.}(2010){Oser}, {Ostriker}, {Naab}, {Johansson}, \&
  {Burkert}}]{Oser2010simu}
{Oser}, L., {Ostriker}, J.~P., {Naab}, T., {Johansson}, P.~H., \& {Burkert}, A.
  2010, \apj, 725, 2312

\bibitem[{{Press} \& {Schechter}(1974)}]{Press_Schechter1974}
{Press}, W.~H. \& {Schechter}, P. 1974, \apj, 187, 425

\bibitem[{{Riechers} {et~al.}(2020){Riechers}, {Hodge}, {Pavesi}, {Daddi},
  {Decarli}, {Ivison}, {Sharon}, {Smail}, {Walter}, {Aravena}, {Capak},
  {Carilli}, {Cox}, {Cunha}, {Dannerbauer}, {Dickinson}, {Neri}, \&
  {Wagg}}]{Riechers2020z5SFR}
{Riechers}, D.~A., {Hodge}, J.~A., {Pavesi}, R., {et~al.} 2020, \apj, 895, 81

\bibitem[{{Roper} {et~al.}(2020){Roper}, {Thomas}, \& {Srisawat}}]{Roper2020}
{Roper}, W.~J., {Thomas}, P.~A., \& {Srisawat}, C. 2020, \mnras, 494, 4509

\bibitem[{{Salim} {et~al.}(2018){Salim}, {Boquien}, \&
  {Lee}}]{Salim2018attenuation}
{Salim}, S., {Boquien}, M., \& {Lee}, J.~C. 2018, \apj, 859, 11

\bibitem[{{Schaye} {et~al.}(2015){Schaye}, {Crain}, {Bower}, {Furlong},
  {Schaller}, {Theuns}, {Dalla Vecchia}, {Frenk}, {McCarthy}, {Helly},
  {Jenkins}, {Rosas-Guevara}, {White}, {Baes}, {Booth}, {Camps}, {Navarro},
  {Qu}, {Rahmati}, {Sawala}, {Thomas}, \& {Trayford}}]{Schaye2015}
{Schaye}, J., {Crain}, R.~A., {Bower}, R.~G., {et~al.} 2015, \mnras, 446, 521

\bibitem[{{Schreiber} {et~al.}(2015){Schreiber}, {Pannella}, {Elbaz},
  {B{\'e}thermin}, {Inami}, {Dickinson}, {Magnelli}, {Wang}, {Aussel}, {Daddi},
  {Juneau}, {Shu}, {Sargent}, {Buat}, {Faber}, {Ferguson}, {Giavalisco},
  {Koekemoer}, {Magdis}, {Morrison}, {Papovich}, {Santini}, \&
  {Scott}}]{Schreiber2015}
{Schreiber}, C., {Pannella}, M., {Elbaz}, D., {et~al.} 2015, \aap, 575, A74

\bibitem[{{Shivaei} {et~al.}(2018){Shivaei}, {Reddy}, {Siana}, {Shapley},
  {Kriek}, {Mobasher}, {Freeman}, {Sanders}, {Coil}, {Price}, {Fetherolf},
  {Azadi}, {Leung}, \& {Zick}}]{Shivaei2018MOSDEF}
{Shivaei}, I., {Reddy}, N.~A., {Siana}, B., {et~al.} 2018, \apj, 855, 42

\bibitem[{{Shuntov} {et~al.}(2022){Shuntov}, {McCracken}, {Gavazzi}, {Laigle},
  {Weaver}, {Davidzon}, {Ilbert}, {Kauffmann}, {Faisst}, {Dubois}, {Koekemoer},
  {Moneti}, {Milvang-Jensen}, {Mobasher}, {Sanders}, \& {Toft}}]{Shuntov2022}
{Shuntov}, M., {McCracken}, H.~J., {Gavazzi}, R., {et~al.} 2022, \aap, 664, A61

\bibitem[{{Sillassen} {et~al.}(2022){Sillassen}, {Jin}, {Magdis}, {Daddi},
  {Weaver}, {Gobat}, {Kokorev}, {Valentino}, {Finoguenov}, {Shuntov},
  {G{\'o}mez-Guijarro}, {Coogan}, {Greve}, {Toft}, \& {Blanquez
  Sese}}]{Sillassen2022}
{Sillassen}, N.~B., {Jin}, S., {Magdis}, G.~E., {et~al.} 2022, \aap, 665, L7

\bibitem[{{Skelton} {et~al.}(2014){Skelton}, {Whitaker}, {Momcheva}, {Brammer},
  {van Dokkum}, {Labb{\'e}}, {Franx}, {van der Wel}, {Bezanson}, {Da Cunha},
  {Fumagalli}, {F{\"o}rster Schreiber}, {Kriek}, {Leja}, {Lundgren}, {Magee},
  {Marchesini}, {Maseda}, {Nelson}, {Oesch}, {Pacifici}, {Patel}, {Price},
  {Rix}, {Tal}, {Wake}, \& {Wuyts}}]{Skelton2014}
{Skelton}, R.~E., {Whitaker}, K.~E., {Momcheva}, I.~G., {et~al.} 2014, \apjs,
  214, 24

\bibitem[{{Somerville} \& {Dav{\'e}}(2015)}]{Somerville_Dave2015}
{Somerville}, R.~S. \& {Dav{\'e}}, R. 2015, \araa, 53, 51

\bibitem[{{Somerville} \& {Primack}(1999)}]{Somerville1999}
{Somerville}, R.~S. \& {Primack}, J.~R. 1999, \mnras, 310, 1087

\bibitem[{{Springel} {et~al.}(2008){Springel}, {Wang}, {Vogelsberger},
  {Ludlow}, {Jenkins}, {Helmi}, {Navarro}, {Frenk}, \&
  {White}}]{Springel2008halo}
{Springel}, V., {Wang}, J., {Vogelsberger}, M., {et~al.} 2008, \mnras, 391,
  1685

\bibitem[{{Springel} {et~al.}(2005){Springel}, {White}, {Jenkins}, {Frenk},
  {Yoshida}, {Gao}, {Navarro}, {Thacker}, {Croton}, {Helly}, {Peacock}, {Cole},
  {Thomas}, {Couchman}, {Evrard}, {Colberg}, \& {Pearce}}]{Springel2005Natur}
{Springel}, V., {White}, S. D.~M., {Jenkins}, A., {et~al.} 2005, \nat, 435, 629

\bibitem[{{Stach} {et~al.}(2019){Stach}, {Dudzevi{\v{c}}i{\={u}}t{\.{e}}},
  {Smail}, {Swinbank}, {Geach}, {Simpson}, {An}, {Almaini}, {Arumugam},
  {Blain}, {Chapman}, {Chen}, {Conselice}, {Cooke}, {Coppin}, {da Cunha},
  {Dunlop}, {Farrah}, {Gullberg}, {Hodge}, {Ivison}, {Kocevski},
  {Micha{\l}owski}, {Miyaji}, {Scott}, {Thomson}, {Wardlow}, {Weiss}, \& {van
  der Werf}}]{Stach2019}
{Stach}, S.~M., {Dudzevi{\v{c}}i{\={u}}t{\.{e}}}, U., {Smail}, I., {et~al.}
  2019, \mnras, 487, 4648

\bibitem[{{Stefanon} {et~al.}(2017){Stefanon}, {Yan}, {Mobasher}, {Barro},
  {Donley}, {Fontana}, {Hemmati}, {Koekemoer}, {Lee}, {Lee}, {Nayyeri}, {Peth},
  {Pforr}, {Salvato}, {Wiklind}, {Wuyts}, {Ashby}, {Castellano}, {Conselice},
  {Cooper}, {Cooray}, {Dolch}, {Ferguson}, {Galametz}, {Giavalisco}, {Guo},
  {Willner}, {Dickinson}, {Faber}, {Fazio}, {Gardner}, {Gawiser}, {Grazian},
  {Grogin}, {Kocevski}, {Koo}, {Lee}, {Lucas}, {McGrath}, {Nandra}, {Newman},
  \& {van der Wel}}]{Stefanon2017EGS}
{Stefanon}, M., {Yan}, H., {Mobasher}, B., {et~al.} 2017, \apjs, 229, 32

\bibitem[{{Taylor}(2005)}]{Taylor2005Topcat}
{Taylor}, M.~B. 2005, in Astronomical Society of the Pacific Conference Series,
  Vol. 347, Astronomical Data Analysis Software and Systems XIV, ed.
  P.~{Shopbell}, M.~{Britton}, \& R.~{Ebert}, 29

\bibitem[{{Trayford} {et~al.}(2015){Trayford}, {Theuns}, {Bower}, {Schaye},
  {Furlong}, {Schaller}, {Frenk}, {Crain}, {Dalla Vecchia}, \&
  {McCarthy}}]{Trayford2015}
{Trayford}, J.~W., {Theuns}, T., {Bower}, R.~G., {et~al.} 2015, \mnras, 452,
  2879

\bibitem[{{Virtanen} {et~al.}(2020){Virtanen}, {Gommers}, {Oliphant},
  {Haberland}, {Reddy}, {Cournapeau}, {Burovski}, {Peterson}, {Weckesser},
  {Bright}, {van der Walt}, {Brett}, {Wilson}, {Jarrod Millman}, {Mayorov},
  {Nelson}, {Jones}, {Kern}, {Larson}, {Carey}, {Polat}, {Feng}, {Moore}, {Vand
  erPlas}, {Laxalde}, {Perktold}, {Cimrman}, {Henriksen}, {Quintero}, {Harris},
  {Archibald}, {Ribeiro}, {Pedregosa}, {van Mulbregt}, \&
  {Contributors}}]{2020SciPy-NMeth}
{Virtanen}, P., {Gommers}, R., {Oliphant}, T.~E., {et~al.} 2020, Nature
  Methods, 17, 261

\bibitem[{{Whitaker} {et~al.}(2011){Whitaker}, {Labb{\'e}}, {van Dokkum},
  {Brammer}, {Kriek}, {Marchesini}, {Quadri}, {Franx}, {Muzzin}, {Williams},
  {Bezanson}, {Illingworth}, {Lee}, {Lundgren}, {Nelson}, {Rudnick}, {Tal}, \&
  {Wake}}]{Whitaker2011}
{Whitaker}, K.~E., {Labb{\'e}}, I., {van Dokkum}, P.~G., {et~al.} 2011, \apj,
  735, 86

\bibitem[{{Wylezalek} {et~al.}(2022){Wylezalek}, {Vayner}, {Rupke}, {Zakamska},
  {Veilleux}, {Ishikawa}, {Bertemes}, {Liu}, {Barrera-Ballesteros}, {Chen},
  {Goulding}, {Greene}, {Hainline}, {Hamann}, {Heckman}, {Johnson}, {Lutz},
  {L{\"u}tzgendorf}, {Mainieri}, {Maiolino}, {Nesvadba}, {Ogle}, \&
  {Sturm}}]{Wylezalek2022jwst}
{Wylezalek}, D., {Vayner}, A., {Rupke}, D. S.~N., {et~al.} 2022, \apjl, 940, L7

\bibitem[{{Zhang} {et~al.}(2008){Zhang}, {Fakhouri}, \&
  {Ma}}]{Zhang2008MergerTree}
{Zhang}, J., {Fakhouri}, O., \& {Ma}, C.-P. 2008, \mnras, 389, 1521

\end{thebibliography}


\begin{appendix}
\section{Comparison of photometric redshifts}

\begin{figure}[ht]
\setlength{\abovecaptionskip}{-0.1cm}
\setlength{\belowcaptionskip}{-0.2cm}
\centering
\includegraphics[width=0.42\textwidth]{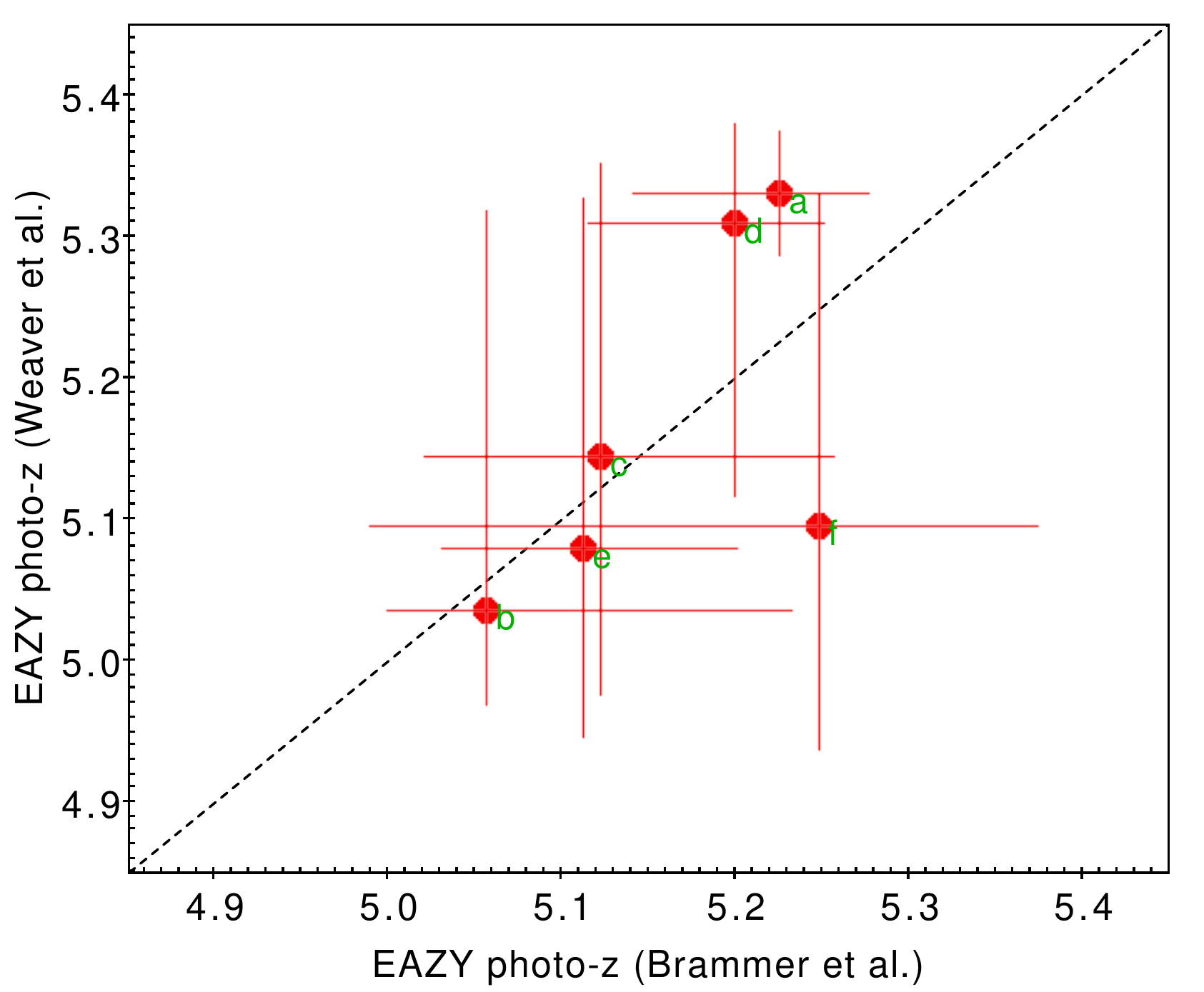}
\includegraphics[width=0.42\textwidth]{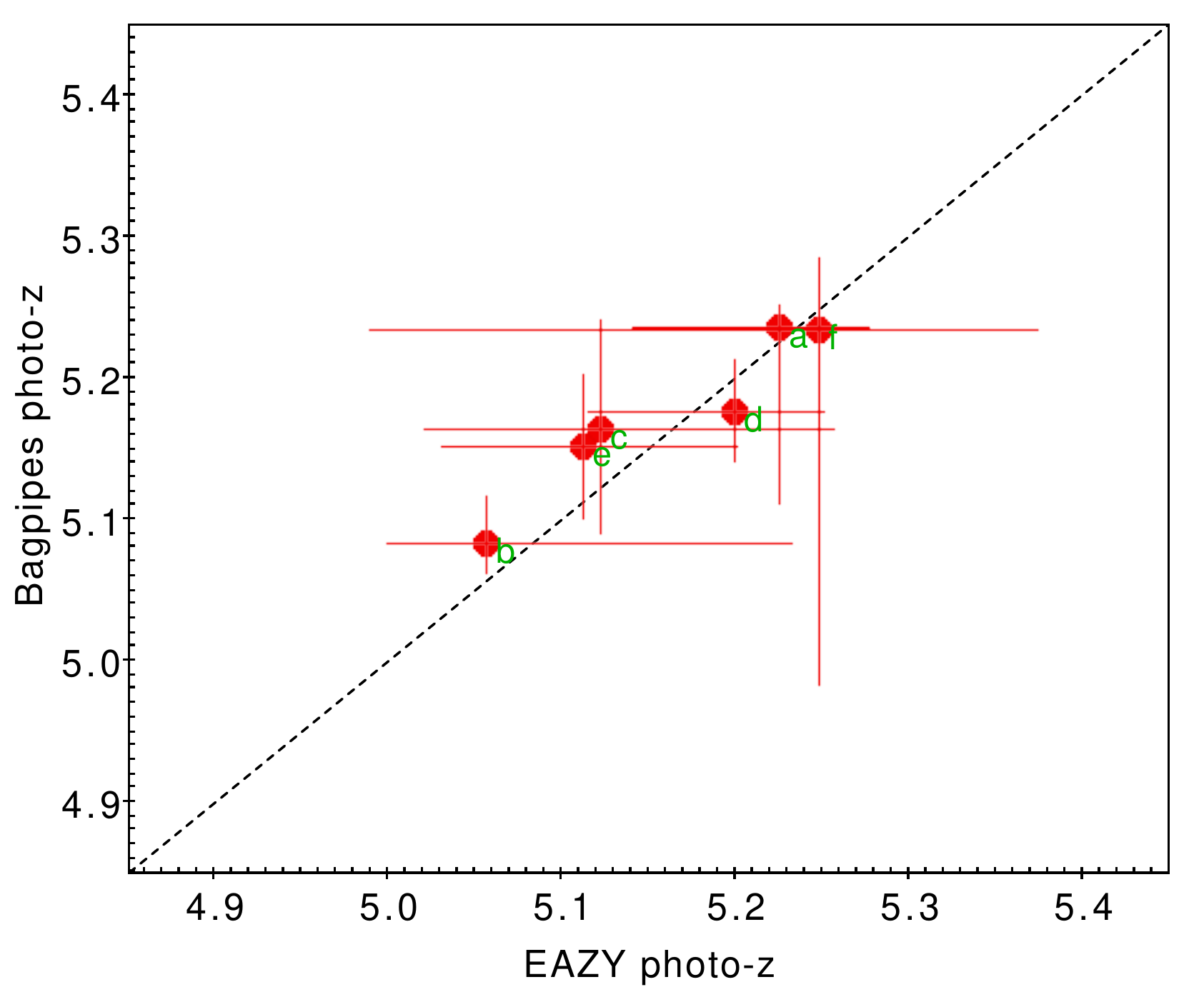}
\caption{
Redshift comparison. {\it Upper:} Comparison of EAZY photo-z computed from two versions of photometry: Weaver et al. and Brammer et al. {\it Bottom:} \textsc{Bagpipes} vs. EAZY photo-$z$, which were all computed using photometry from Brammer et al. Group members are labeled with green text.
\label{z_compare}
}
\end{figure}

\begin{figure}[ht]
\setlength{\abovecaptionskip}{-0.1cm}
\setlength{\belowcaptionskip}{-0.2cm}
\centering
\includegraphics[width=0.49\textwidth]{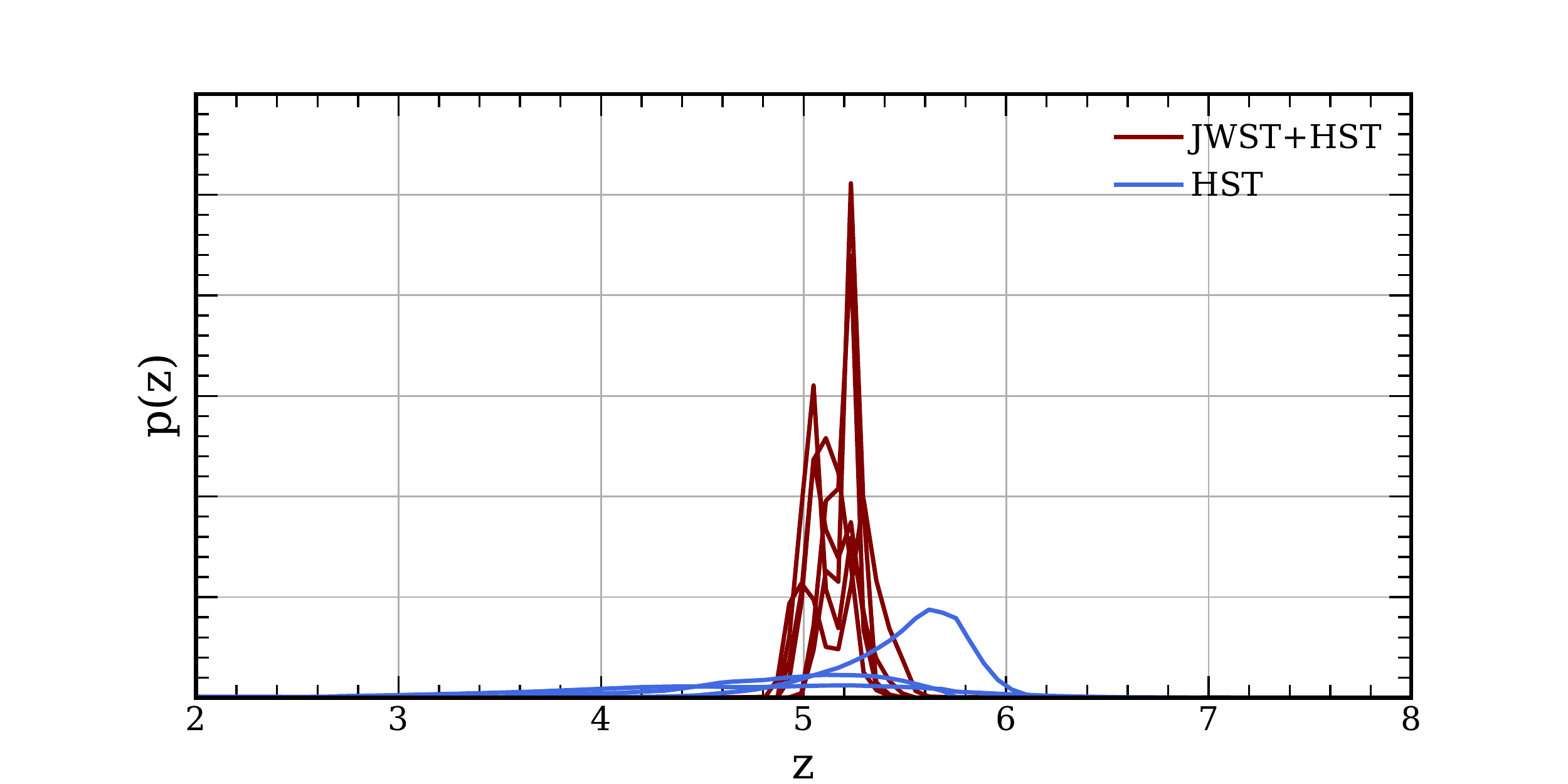}
\caption{%
Comparison of PDF(z) with and without JWST, normalized to the integral. JWST+HST PDF(z) are shown for the six candidate members, while HST PDF(z) are only shown for the three HST/F160W-detected sources.  
\label{zPDF}
}
\end{figure}

We present the comparison of photometric redshifts computed from different catalogs and SED codes in Fig.~\ref{z_compare}. In Fig.~\ref{zPDF}, we compare the photo$-z$ PDF(z) with and without JWST data.

\end{appendix}

\end{document}